\documentclass[times]{aastex62}

\usepackage{acronym}
\newacro{CDM}{cold dark matter}
\newcommand{\CDM}{\ac{CDM}}
\newacro{FDM}{fuzzy dark matter}
\newcommand{\FDM}{\ac{FDM}}
\newacro{DM}{dark matter}

\newacro{DF}{distribution function}
\newcommand{\DF}{\ac{DF}}

\usepackage{commath}
\usepackage{bm}
\usepackage{mathtools}
\DeclareMathOperator{\erf}{erf}
\renewcommand{\inf}{\infty}

\newcommand{\rd}{\mathrm{d}}
\newcommand{\rb}{\mathrm{b}}
\newcommand{\re}{\mathrm{e}}
\newcommand{\ri}{\mathrm{i}}
\newcommand{\rt}{\mathrm{t}}
\newcommand{\rf}{\mathrm{f}}
\newcommand{\rr}{\mathrm{r}}
\newcommand{\rc}{\mathrm{c}}

\newcommand{\rp}{\mathrm{p}}
\newcommand{\rh}{\mathrm{h}}
\newcommand{\mb}{m_\rb}

\newcommand{\pc}{\,\mathrm{pc}}
\newcommand{\kpc}{\,\mathrm{kpc}}
\newcommand{\eV}{\,\mathrm{eV}}
\newcommand{\kms}{\,\mathrm{km\ s}^{-1}}

\newcommand{\Gyr}{\,\mathrm{Gyr}}
\newcommand{\rs}{\mathrm{s}}

\newcommand{\ccdot}{\!\cdot\!}
\newcommand{\clog}{\!\log}
\usepackage{xargs}
\newcommandx\dint[2][usedefault, addprefix=\global, 1=,2=]{\int_{#1}^{#2}\!\! \rd}

\newcommand\cpar[1]{\left(\! #1 \!\right)}
\newcommand\power[2]{{\cpar{#1}}^{\!\! #2}}

\newcommand\eq{equation}
\newcommand\eqs{equations}
\newcommand\Eq{Equation}
\newcommand\Eqs{Equations}

\newcommand{\bx}{\mathbf{x}}
\renewcommand{\bv}{\mathbf{v}}
\newcommand{\bk}{\mathbf{k}}

\newcommand{\br}{\mathbf{r}}
\newcommand{\bF}{\mathbf{F}}

\newcommand{\vdmin}{v_{d,\min}}

\newcommand{\kmin}{k_{\min}}
\newcommand{\kmax}{k_{\max}}
\newcommand{\meff}{m_\mathrm{eff}}
\newcommand{\mueff}{\mu_\mathrm{eff}}
\newcommand{\seff}{\sigma_\mathrm{eff}}
\newcommand{\Xeff}{X_\mathrm{eff}}
\newcommand{\hPhi}{\widehat{\Phi}}
\newcommand{\CPhi}{C_{\Phi}}
\newcommand{\hCPhi}{\widehat{C}_{\Phi}}
\newcommand{\hrho}{\widehat{\rho}}
\newcommand{\Crho}{C_{\!\rho}}
\newcommand{\hCrho}{\widehat{C}_{\!\rho}}
\newcommand{\hW}{\widehat{W}}
\newcommand{\hk}{\widehat{k}}
\newcommand{\hbk}{\widehat{\mathbf{k}}}

\newcommand{\brp}{\mathbf{r}^{\prime}}
\newcommand{\tp}{t^{\prime}}
\newcommand{\bkp}{\mathbf{k}^{\prime}}
\newcommand{\bkpp}{\mathbf{k}''}
\newcommand{\omegap}{\omega^{\prime}}
\newcommand{\bvp}{\mathbf{v}^{\prime}}
\newcommand{\brpp}{\mathbf{r}^{\prime\prime}}
\newcommand{\kp}{k^{\prime}}
\newcommand{\kpp}{k^{\prime\prime}}
\newcommand{\sprime}{s^{\prime}}
\newcommand{\vp}{v^{\prime}}
\newcommand{\bvd}{\mathbf{v}_{\mathrm{d}}}
\newcommand{\hbvd}{\widehat{\mathbf{v}}_{\mathrm{d}}}
\newcommand{\Tc}{T_{\mathrm{c}}}
\newcommand{\bvc}{\mathbf{v}_{\mathrm{c}}}
\newcommand{\veld}{v_{\mathrm{d}}}
\newcommand{\velc}{v_{\mathrm{c}}}
\newcommand{\veps}{\varepsilon}

\newcommand{\rhob}{\rho_{\rb}}
\newcommand{\rhop}{\rho_\mathrm{p}}
\newcommand{\Fb}{F_{\rb}}
\newcommand{\Fp}{F_{\mathrm{p}}}
\newcommand{\mpp}{m_{\mathrm{p}}}
\newcommand{\oFb}{F_{\mathrm{eff}}}

\newcommand{\rri}{r_{\mathrm{i}}}

\newcommand{\bmax}{b_{\max}}
\newcommand{\bmin}{b_{\min}}
\newcommand{\lsig}{\lambda_\sigma}
\newcommand{\lbarsig}{\lambdabar_\sigma}

\usepackage{soul}

\shorttitle{fuzzy dark matter}
\shortauthors{Bar-Or, Fouvry and Tremaine}
\begin{document}

\title{Relaxation in a Fuzzy Dark Matter Halo}

\author[0000-0002-8927-4571]{Ben Bar-Or}
\affiliation{Institute for Advanced Study, Princeton, NJ 08540, USA}

\author[0000-0002-0030-371X]{Jean-Baptiste Fouvry}
\altaffiliation{Hubble Fellow}
\affiliation{Institute for Advanced Study, Princeton, NJ 08540, USA}

\author[0000-0002-0278-7180]{Scott Tremaine}
\affiliation{Institute for Advanced Study, Princeton, NJ 08540, USA}

\acused{FDM} \begin{abstract}
  Dark matter may be composed of light bosons, ${\mb \sim 10^{-22} \eV}$, with
  a de~Broglie wavelength $\lambda \sim 1 \kpc$ in typical galactic
  potentials. Such ``fuzzy'' dark matter (\FDM) behaves like \CDM\ on much
  larger scales than the de~Broglie wavelength, but may resolve some of the
  challenges faced by \CDM\ in explaining the properties of galaxies on small
  scales ($\lesssim 10 \kpc$). Because of its wave nature, \FDM\ exhibits
  stochastic density fluctuations on the scale of the de~Broglie wavelength
  that never damp. The gravitational field from these fluctuations scatters
  stars and black holes, causing their orbits to diffuse through phase
  space. We show that this relaxation process can be analyzed quantitatively
  with the same tools used to analyze classical two-body relaxation in an
  $N$-body system, and can be described by treating the \FDM\ fluctuations as
  quasiparticles, with effective mass
  $\sim 10^7 M_\odot {(1\kpc/r)}^2{(10^{-22}\eV/\mb)}^3$ in a galaxy with a
  constant circular speed of $200\kms$. This novel relaxation mechanism may
  stall the inspiral of supermassive black holes or globular clusters due to
  dynamical friction at radii of a few hundred parsecs, and can heat and expand the
  central regions of galaxies. These processes can be used to constrain the
  mass of the light bosons that might comprise \FDM\@.
\end{abstract}


\fontdimen16\textfont2 = \fontdimen17\textfont2

\acresetall{}

\section{Introduction} 
\label{sec:intro}

Despite the remarkable success of cosmological models based on \CDM\ in
explaining large-scale structure and other cosmological phenomena, \CDM\ has
faced challenges in predicting aspects of small-scale structure, such as the
abundance of dwarf galaxies and the dark-matter density near the centers of
galaxies~\citep[e.g.,][]{Weinberg+2015}. The solution to these problems may lie
either in baryonic physics (e.g., feedback to the interstellar medium from
supernovae or black holes) or in the properties of the dark matter itself.

In this paper, we examine aspects of the behavior of \FDM, which is composed of
bosons with extremely small masses, typically
${\mb\sim 10^{-21}\mbox{--}10^{-22} \eV}$~\citep[e.g.,][]{Hu+2000,
  Hui+2017}. The corresponding de~Broglie wavelength at velocity $v$ is
$\lambda = h/(\mb v)\simeq 1.20\kpc\, (10^{-22}\eV /\mb)\allowbreak(100
\kms/v)$; on scales much larger than this, \FDM\ behaves like \CDM, but on small
scales, it exhibits quite different properties and may match the observations
better than \CDM~\citep[e.g.,][]{Hui+2017}.

Several studies have argued that the mass range
${\mb\lesssim\mbox{\,a few}\times 10^{-21}\eV}$ is ruled out by constraints
from the Lyman-$\alpha$ forest power spectrum~\citep{viel+2013,Armengaud+2017,
  Irsic+2017,kobayashi+2017,Nori+2019}, which would imply that \FDM\ is
indistinguishable from \CDM\ in its effects on observed small-scale
structure. However, (i) these constraints usually rely on assumptions (e.g.,
uniform ionizing background) that are plausible but may be oversimplified, (ii)
the mass constraint can be much weaker in variants of the \FDM\ model
\citep[e.g.,][]{leong+2018}, and (iii) the dynamical processes discussed here can
be important whether or not the \FDM\ particle mass is small enough to
influence small-scale structure.

Since \FDM\ particles are bosons, density fluctuations in the dark matter are
correlated over a distance of the order of the de~Broglie
wavelength.~\citet{Hui+2017} argued that the fluctuating gravitational force
from an \FDM\ field of mean density $\rho$ is similar to that of a classical
$N$-body system composed of quasiparticles with effective mass
${ \meff \sim \rho \lambda^{3} }$.

The relaxation time of a test particle orbiting in a stellar system of radius
$R$, containing $N$ bodies of mass $m$, is~\citep{Binney+2008}
\begin{equation}
  \label{eq:intro}
  t_\rr \sim \frac{\sigma^3R^3}{G^2m^2 N \log N} \sim \frac{N t_\rd}{\log N },
\end{equation}
where $ t_\rd \sim R/\sigma $ is the dynamical time and the typical velocity
$\sigma\sim {(G m N/R)}^{1/2}$ by the virial theorem. In plausible \CDM\
models, the particle mass $m$ is so small that the relaxation time is much
larger than the age of the universe, so dark halos are collisionless. In \FDM\
models, however, the effective mass $\meff$ of the quasiparticles is large
enough that relaxation can be important. Relaxation between the quasiparticles
leads to the formation and growth of a central Bose--Einstein condensate or
soliton, a process that we do not study here. Relaxation between the
quasiparticles and macroscopic objects such as stars can heat, and therefore
expand, the stellar system as it evolves toward equipartition with the
quasiparticles. Relaxation between the quasiparticles and massive objects such
as black holes or globular clusters can stall the inspiral of the massive
object toward the center of the galaxy that is otherwise caused by dynamical
friction from both baryonic and dark matter.

The goal of this paper is to place these physical arguments on a firm
quantitative basis by analyzing the nature of the relaxation of classical
particles, both test particles and massive objects, in an infinite \FDM\ halo
that is homogeneous in the mean. The paper is organized as follows. In
Section~\ref{sec:dcs}, we derive the velocity diffusion coefficients for a
zero-mass test particle traveling at constant velocity in a homogeneous halo
with stochastic density fluctuations described by a correlation function. These
results are used in Section~\ref{sec:tb} to re-derive the classical formulae
for the diffusion coefficients in a gravitational $N$-body system and in
Section~\ref{sec:dc_fdm} to derive the analogous formulae for an \FDM\ halo. We
find that there are remarkable parallels between the results of the two
calculations. In Section~\ref{sec:dyf}, we consider the steady force acting on a
massive object traveling through an \FDM\ halo (dynamical friction), which
completes the standard set of diffusion coefficients. We then turn in
Section~\ref{sec:ms} to a brief discussion of two applications: the inspiral of
a massive object into the galactic center by dynamical friction, which can be
halted by relaxation at scales where the mass of the object becomes comparable
to the mass of the \FDM\ quasiparticles, and the expansion and heating of a
stellar system such as a bulge embedded in an \FDM\ halo. We summarize and
conclude in Section~\ref{sec:summary}.

\subsection{The Coulomb logarithm}

\label{sec:coulomb}

The factor $\log N$ in \eq~\eqref{eq:intro} is known as the Coulomb
logarithm~\citep{Binney+2008}. More generally, it is written as $\log\Lambda$,
where $\Lambda \equiv \bmax / \bmin$ is the ratio between the maximum and
minimum scales of the encounters that dominate the diffusion coefficients or
relaxation time.

The calculations in this paper are for an infinite homogeneous system, for
which we invoke the Jeans swindle, that is, we neglect any acceleration due to the
homogeneous average density~\citep{Binney+2008}. This assumption breaks down on
scales larger than the Jeans length ${(\sigma^2/ G \rho)}^{1/2}$, which for a
finite system is comparable to the system's radius $R$ by the virial
theorem. At scales much larger than $R$, the density is negligible so
$\bmax\lesssim R$. In addition, if the system is centrally concentrated and the
orbital size $r\ll R$, then the effects of encounters on a scale $b$ with
$r\ll b\ll R$ will average to zero.\footnote{This is not true for systems in
  which there is a global resonance, such as spherical or nearly Keplerian
  systems~\citep[e.g.,][]{Rauch+1996,Kocsis+2015}. In such cases, ``resonant
  relaxation'' implies that the slow action associated with the global
  resonance relaxes much faster than the other actions.} Thus, we can set
$\bmax \simeq \min(r,R)$.

In classical $N$-body systems composed of point particles of mass $\mpp$, we
set $\bmin\simeq b_{90}\equiv G\mpp/\sigma^2$; this is the impact parameter at
which a typical particle is deflected by $90^\circ$. If the particles have a
non-zero scale length $\veps$, either because they represent star clusters or
other sub-systems of non-zero size, or because they have been artificially
``softened'' to reduce integration errors, we set $\bmin\simeq \veps$ (see
Section~\ref{sec:dcs} for more details). Finally, in modeling relaxation due to
\FDM\ quasiparticles we will show that it is appropriate to set the minimum
scale to half of the typical de~Broglie angular wavelength,\footnote{We
  distinguish between the typical de~Broglie wavelength $\lsig = h/(\mb\sigma)$
  and the typical de~Broglie \emph{angular} wavelength
  $\lbarsig = \hbar/(\mb\sigma)$.}
$\bmin \simeq \lbarsig/2 \equiv \hbar/(2\mb \sigma)$.

These considerations lead us to define classical, softened, and \FDM\ Coulomb
factors:
\begin{equation}
  \label{eq:coulomb}
  \Lambda_\mathrm{cl}\equiv \frac{\bmax}{b_{90}}, 
  \quad \Lambda_\mathrm{soft}\equiv \frac{\bmax}{\veps}, 
  \quad \Lambda_\mathrm{FDM}\equiv \frac{2\bmax}{\lbarsig}.
\end{equation}
The precise value of $\Lambda$ is uncertain by a factor of order unity because
$\bmax$ cannot be determined exactly from calculations in an infinite
homogeneous medium.\footnote{Accurate treatments of the diffusion coefficients
  in inhomogeneous equilibrium stellar systems require the analysis of orbital
  resonances~\citep{Tremaine+1984,Binney+1988, Heyvaerts2010, Chavanis2012,
    Fouvry+2018}.} This ambiguity is only a minor concern in the common case
where $\Lambda\gg 1$. Therefore, for clarity, we shall always express our
formulae in terms of one of the three quantities in \eq~\eqref{eq:coulomb},
even when the derivation yields a value for the argument of the log that
differs from these by a factor of order unity.

\section{Diffusion of a test particle in a fluctuating density field}
\label{sec:dcs}

In this section, we calculate the velocity diffusion coefficients for a
zero-mass test particle embedded in a potential that exhibits stochastic
fluctuations but is on average uniform in space and stationary in time. The
stochastic fluctuations drive the evolution of the test particle's velocity, and
their spatial and temporal correlations determine the degree to which the test
particle can respond to these fluctuations.\footnote{A similar approach was
  pioneered by~\cite{Cohen1975} who calculated the temporal and spatial
  correlations of the stochastic forces in a finite homogeneous stellar
  system.}  Here we also restrict ourselves to the linear approximation, which
for classical particles is equivalent to assuming that $\veps\gg b_{90}$ in the
notation of Section~\ref{sec:coulomb}. This approximation is valid for most
cases of interest involving \FDM\@.

Consider a time-dependent potential $\Phi(\br,t)$ with zero mean,
$\langle \Phi(\br,t)\rangle=0$, and stationary correlation function,
\begin{equation}
  \label{eq:Phi_corr}
  \langle \Phi(\br, t) \, \Phi(\brp, \tp)  \rangle = \CPhi(\br - \brp, t - \tp);
\end{equation}
here ${ \langle \,\cdot\, \rangle }$ denotes an ensemble average, i.e., an
average over all possible realizations of the potential. It is useful to write
the potential in terms of its temporal and spatial Fourier transform,
${ \hPhi (\bk, \omega) }$, defined by
\begin{equation}
  \label{eq:Phi_ft}
  \Phi(\br, t) = \!\iint\!\frac{\rd \bk \rd \omega}{{(2\pi)}^4}
  \, \hPhi (\bk, \omega) \, \re^{\ri \bk\cdot\br - \ri \omega t}.
\end{equation}
The correlation function of ${ \hPhi (\bk, \omega) }$ is given by
\begin{equation}
  \label{eq:Phi_ft_corr}
  \langle \hPhi (\bk, \omega) \, \hPhi^{*}(\bkp , \omegap) \rangle =
  {(2\pi)}^4 \hCPhi (\bk, \omega) \delta (\omega - \omegap) \delta(\bk - \bkp),
\end{equation}
where ${ \hCPhi (\bk, \omega) }$ is the temporal and spatial Fourier transform
of the potential correlation function ${ \CPhi(\br, t) }$.

Given the potential in \eq~\eqref{eq:Phi_ft}, the acceleration of the test
particle is
\begin{equation}
  \label{eq:eom}
  \dot{\bv}(\br, t) = -\nabla \Phi(\br,t) = 
  - \ri \!\iint\frac{\bk \rd \bk \rd \omega}{{(2\pi)}^4} \,
  \hPhi (\bk, \omega) \, \re^{\ri \bk\cdot\br - \ri \omega t},
\end{equation}
and its change in velocity over time $t$ is given by
\begin{equation}
  \label{eq:dv}
  \Delta \bv(t) = \dint[0][t] s\, \dot{\bv}[\br(s), s].
\end{equation}
As we assume that the mean force is zero, we can expand ${ \br(t) }$ around the
initial position and velocity,
\begin{equation}
  \label{eq:rt}
  \br(t) = \br_0 + \bv_0 t + \dint[0][t] s \, (t-s) \, \dot{\bv}(\br_0+\bv_0 s,s) + \cdots .
\end{equation}
Thus, the change in velocity is given by
\begin{align}
  \label{eq:dv_exp}
  \Delta \bv(t) = {}
  &
    -\ri \!\iint\frac{\bk \rd \bk \rd \omega}{{(2\pi)}^4}
    \hPhi (\bk, \omega) \, \re^{\ri \bk\cdot\br_0} \dint[0][t] s \,
    \re^{\ri(\bk\cdot\bv_0 - \omega) s}
    \nonumber \\
  & + \ri\!\iint\frac{\bk \rd \bk \rd \omega}{{(2\pi)}^4}
    \!\!\iint\frac{\bk \ccdot \bkp \rd \bkp \rd \omegap}{{(2\pi)}^4}
    \, \hPhi (\bk, \omega) \, \hPhi^{*} (\bkp, \omegap) 
    \re^{\ri (\bk - \bkp) \cdot\br_0}\dint[0][t] s \, \re^{\ri(\bk\cdot\bv_0-\omega) s}    
    \dint[0][s] \sprime \, (s - \sprime) \, \re^{-\ri(\bkp \cdot \bv_0-\omegap ) \sprime},
\end{align}
in which we have kept only terms up to second order in
${ \hPhi (\bk, \omega) }$.

To proceed forward, we assume that changes in velocity result from the
accumulation of many small increments. As a result, the velocity evolution can
be described by a Fokker--Planck equation in which the first and second
diffusion coefficients are the first and second moments of the transition
probability, namely ${ D[\Delta v_i] = \langle \Delta v_i (T) \rangle/T }$ and
${ D[\Delta v_i \Delta v_j] = \langle \Delta v_i(T) \, \Delta v_j(T) \rangle/T
}$, where ${\Delta v_i(T)}$ is the change in velocity component $i$ over a time
$T$. This is equivalent to ignoring higher moments of the transition
probability~\citep[e.g.,][]{Henon1960,Risken1989} and is usually a good
assumption if the first and second moments of the transition probability are
finite.

Therefore, the diffusion coefficients are given by\footnote{In deriving the
  first of these equations, we have made use of the fact that $\Phi(\br,t)$ is
  real, so $\hCPhi(-\bk,-\omega)=\hCPhi(\bk,\omega)$ for real $\omega$ and
  $\bk$.}
\begin{equation}
  \label{eq:Dv}
  D[\Delta v_i] = \frac{1}{2} \sum_j \dpd{}{v_j}\,
  \!\!\iint\! 
  \frac{ \rd \bk \rd \omega}{{(2\pi)}^3}k_i k_j
  \, \hCPhi (\bk, \omega) 
  \, K_T (\omega-\bk\cdot\bv),
\end{equation}
and
\begin{equation}
  \label{eq:Dvv}
  D[\Delta v_i \Delta v_j] = 
  \!\!\iint\! 
  \frac{\rd \bk \rd \omega}{{(2\pi)}^3} k_i k_j
  \hCPhi (\bk, \omega) K_T(\omega-\bk\cdot\bv),
\end{equation}
where 
\begin{equation}
  \label{eq:KT}
  K_T(\omega) \equiv \frac{1}{2\pi T}
  \!\!\int_0^T\! \!\int_0^T \! \rd s \rd \sprime
  \re^{\ri \omega (s - \sprime)} = \frac{1-\cos(\omega T)}{\pi \omega^2 T},
\end{equation}
is the finite-time kernel, which is normalized such that
${ \! \int \! \rd \omega \, K_T(\omega) = 1 }$. In the limit ${ T\to \inf }$,
$K_T(\omega) \to \delta(\omega)$ and $D[\Delta v_i \Delta v_j]$ becomes
time-independent, and the process is diffusive. On short timescales,
$K_{T} \to T/(2\pi)$, the process is ballistic, and
${D[\Delta v_i \Delta v_j] \simeq \langle \dot{v}_i\dot{v}_j \rangle T}$
describes the instantaneous coherent (in time) force acting on the test
particle.

\Eqs~\eqref{eq:Dv} and~\eqref{eq:Dvv} satisfy the relation
\begin{equation}
  \label{eq:Dv_fd}
  D[\Delta v_i] =  \frac{1}{2}\sum_j 
  \dpd{}{v_j} D[\Delta v_i \Delta v_j],
\end{equation}
which is the fluctuation-dissipation relation for a zero-mass test
particle~\citep[e.g.,][]{Binney+1988,Binney+2008}.\footnote{This derivation is more general
  than the one in~\citet{Binney+2008}, which requires that the distribution
  function of the particles that cause the potential fluctuations is isotropic
  in velocity space.}

We now assume that there is a finite correlation time $\Tc$ such that
${ \CPhi(\br, t)\to 0 }$ for ${ |t|\gg \Tc }$, and that this correlation time
is much shorter than any other time of interest. This assumption allows us to
take the limit ${ T \to \infty }$, in which the kernel ${ K_T(\omega) }$ can be
approximated as a delta function. \Eqs~\eqref{eq:Dv} and~\eqref{eq:Dvv} then
read
\begin{align}
  \label{eq:DvII}
  D_{i} & = D[\Delta v_i] =  \frac{1}{2}\sum_j \dpd{}{v_j} \!\int\! \frac{\rd \bk}{{(2\pi)}^3} k_i k_j
          \hCPhi (\bk, \bk\ccdot\bv),\\
  \label{eq:DvvII}
  D_{ij} & = D[\Delta v_i \Delta v_j] = \!\!\int\!\ \frac{\rd \bk}{{(2\pi)}^3} k_i k_j
           \hCPhi (\bk, \bk\ccdot\bv).
\end{align}

Under these approximations, the probability distribution of the velocity of a
test particle, ${ P (\bv,t) }$, is governed by the Fokker--Planck equation
\begin{equation}
  \label{eq:FP}
  \dpd{P (\bv,t)}{t} = -\sum_i\dpd{}{v_i}\big[D_i \, P(\bv,t)\big]
   +\frac{1}{2}\sum_{ij}\dpd{^2}{v_i\partial v_j}\big[D_{ij} \, P (\bv,t)\big]
  =\frac{1}{2}\sum_{ij}\dpd{}{v_i}\!
  \bigg[
  \!
  D_{ij} \, 
  \!
  \dpd{P (\bv,t)}{v_j}
  \bigg],
\end{equation}
where the last equality is derived using \eq~\eqref{eq:Dv_fd}.

We now specialize to the case in which the potential fluctuations arise from
density fluctuations $\rho (\br, t)$ around a mean field density $\rhop$, so
${\langle \rho(\br, t) \rangle = 0}$. Assuming that these fluctuations are a
stationary homogeneous random field, the correlation function of the density
fluctuations can be written as
\begin{equation}
  \label{eq:rho_corr}
  \langle \rho(\br, t) \, \rho(\brp , \tp) \rangle = C_\rho(\br - \brp, t - \tp).
\end{equation}
The potential fluctuations associated with the density fluctuations
$\rho(\br, t)$ are given by the Fourier transform of Poisson's equation,
${ \hPhi (\bk ,\omega) = - 4 \pi G \hrho (\bk, \omega)/k^2 }$, $k = |\bk|$, and the Fourier
transforms of the correlation functions are related by
\begin{equation}
  \label{eq:C_R}
 \hCPhi(\bk, \omega) = \frac{16 \pi^2 G^2}{k^4} \hCrho (\bk, \omega).
\end{equation}
The diffusion coefficients, \eqs~\eqref{eq:DvII} and~\eqref{eq:DvvII}, become
\begin{align}
\label{eq:Dv_rho}
 D_i &= 
    \frac{G^2}{\pi} \sum_j \dpd{}{v_j}\, \dint \bk \frac{k_i k_j}{k^4}
    \hCrho (\bk, \bk\ccdot\bv),\\
\label{eq:Dvv_rho}
 D_{ij} &= 
    \frac{2G^2}{\pi} \dint \bk \frac{k_i k_j}{k^4}
    \hCrho (\bk, \bk\ccdot\bv).
\end{align}

\subsection{Classical two-body relaxation}
\label{sec:tb}

We now use the results from the preceding discussion to obtain the diffusion
coefficients for a zero-mass test particle interacting with an infinite,
homogeneous system of classical ``field'' particles of individual mass $\mpp$,
characterized by a \DF\ ${ \Fp (\bv) }$. Here, the \DF\ is normalized such that
${ \!\int\! \rd \bv \Fp (\bv) }$ is the mass density $\rhop$, and we ignore the
self-gravity of the particles, considering only the gravitational forces that
they exert on the test particle. This is the system examined in the classic
work of~\citet{Chandrasekhar1942,Chandrasekhar1943}.

Given our assumptions, each field particle travels on a straight line at
constant velocity. Then, the density fluctuations around the mean density
$\rhop$ are given by
\begin{equation}
  \label{eq:rho_nb}
  \rho(\br,t) = \mpp \sum_{n} \delta (\br - \br_n - \bv_n t) - \rhop ,
\end{equation}
where ${ (\br_n, \bv_n )}$ stands for the position and velocity of the field particle $n$
at time $t=0$. The associated density correlation function is
\begin{equation}
  \label{eq:rho_c}
  \Crho(\br-\brp,t-\tp)=\langle \rho(\br,t) \, \rho(\brp , \tp) \rangle
         = \mpp \dint \bv\, \delta[\br - \brp - \bv (t - \tp )] \Fp (\bv) ,
\end{equation}
and its temporal and spatial Fourier transform is
\begin{equation}
  \label{eq:Bft}
  \hCrho (\bk,\omega) = 2\pi \mpp \dint \bv'\, \delta(\bk\cdot\bv'
  -\omega) \Fp(\bv').
\end{equation}

From \eqs~\eqref{eq:Dv_rho} and~\eqref{eq:Dvv_rho}, we obtain the first and
second-order diffusion coefficients,
\begin{align}
  \label{eq:Dv_2b}
  D_i = {}
  &
    2 G^2\mpp \clog\Lambda  \dint\hbk \,\hk_i 
    \dint \bvp   \,
    \hbk \ccdot \dpd{}{\bv} 
    \delta[\hbk \ccdot (\bvp - \bv) ] \Fp(\bv')
    \nonumber \\
  = {}
  &
    2 G^2\mpp \clog\Lambda  \dpd{}{v_i} \dint\hbk \dint \bvp\,
    \delta[\hbk \ccdot (\bvp - \bv) ] \Fp(\bv')
    \nonumber \\
  = {}
  &
    4\pi G^2\mpp \clog\Lambda \,\dpd{}{v_i} 
    \! \int \! \rd \bvp
    \frac{\Fp (\bvp)}{|\bv - \bvp|},
\end{align}
and
\begin{align}
  \label{eq:Dvv_2b}
  D_{ij} =  {}
  &
    4 G^2 \mpp \,\clog\Lambda
    \dint \hbk \, \hk_i \hk_j\dint \bvp \,
    \delta[\hbk \ccdot (\bvp - \bv) ]\Fp (\bvp) 
    \nonumber \\
  = {}
  &
    2 G^2 \mpp \,\clog\Lambda
    \frac{\partial^2}{\partial v_i \partial v_j} 
    \dint \hbk \, \dint \bvp \,
    |\hbk \ccdot (\bvp - \bv)|\Fp (\bvp) 
    \nonumber \\
  = {}
  &
    4\pi G^2 \mpp \clog\Lambda \frac{\partial^2}{\partial v_i \partial v_j} 
    \int \rd \bvp\,
    |\bv-\bvp| \Fp (\bvp),
\end{align}
where $\hbk\equiv \bk/|\bk|$. Here,
${ \log\Lambda = \!\int\! \rd k / k =\log\kmax/\kmin}$ is the Coulomb logarithm
(Section~\ref{sec:coulomb}), because we can identify $1/\kmin$ and $1/\kmax$ as
the maximum and minimum scales $\bmax$ and $\bmin$ to within factors of order
unity. To obtain \eqs~\eqref{eq:Dv_2b} and~\eqref{eq:Dvv_2b}, we used the
relation
\begin{equation}
 \label{eq:delta_int}
 \dpd{}{x_{j_1}}\dots \dpd{}{x_{j_\ell}} \dint \hbk\, \hk_{i_1}\dots\hk_{i_n}\delta(\hbk\ccdot\bx)
 = \dpd{}{x_{i_1}}\dots \dpd{}{x_{i_n}} \dint \hbk \,  \hk_{j_1}\dots\hk_{j_\ell}\Delta_{n-\ell}(\hbk\ccdot\bx),
\end{equation}
where
\begin{equation}
  \label{eq:Delta_n}
     \Delta_{n}(x) = 
   \begin{cases}
  \displaystyle \frac{1}{2(n-1)!}
    \frac{x^n}{|x|}, & n > 0,
   \\
    \delta^{(n)}(x), & n \le 0,
 \end{cases}
\end{equation}
is the $n$th integral (derivative) of the Dirac delta function $\delta(x)$.

The diffusion coefficients in \eqs~\eqref{eq:Dv_2b} and~\eqref{eq:Dvv_2b} are
identical to the standard diffusion
coefficients~\citetext{\citealt{Rosenbluth+1957}; see
also~\citealt{Chavanis2013b}} for a zero-mass test particle in
an infinite homogeneous medium, up to the usual ambiguity in the precise
definition of the Coulomb logarithm.

Plugging the diffusion coefficients into the Fokker-Planck
equation~\eqref{eq:FP}, we obtain the (homogeneous) Landau equation~\citep{Landau1936} for a
zero-mass test particle,
\begin{equation}
  \label{eq:Landau}
  \dpd{P (\bv, t)}{t} = 
    2G^2 \mpp  \sum_{ij} \dpd{}{v_i} \dint \bk \dint \bv' \,
    \frac{k_i k_j}{k^4} 
    \delta [\bk \ccdot (\bv - \bv')]
     \Fp(\bv') 
     \! \dpd{}{v_j}  \!\!   P (\bv, t).
\end{equation}
See~\citet{Chavanis2013b} for the historical connection between this
equation and the later treatments of~\citet{Chandrasekhar1942,
  Chandrasekhar1943} and~\citet{Rosenbluth+1957}.

For a Maxwellian velocity distribution,
\begin{equation}
\label{eq:maxwell}
\Fp(\bv)=\frac{\rhop}{{(2\pi\sigma^2)}^{3/2}} \, \re^{ - v^{2} / (2 \sigma^{2})},
\end{equation}
with $v = |\bv|$; the integral expressions for the diffusion coefficients, \eqs~\eqref{eq:Dv_2b}
and~\eqref{eq:Dvv_2b}, can be evaluated explicitly. The diffusion coefficients
in the directions parallel and perpendicular to the test-particle velocity
are~\citep{Binney+2008}
\begin{align}
  \label{eq:Dpar}
  D[\Delta v_\parallel] = & -\frac{4\pi G^2 \rhop \mpp \log\Lambda}{\sigma^2} \mathbb{G}(X) \\
  \label{eq:D2par}
  D[(\Delta v_\parallel^2)] = & \frac{4 \sqrt{2} \pi G^2 \rhop \mpp
    \log\Lambda}{\sigma}
  \frac{\mathbb{G}(X)}{X}, \\
  \label{eq:D2per}
  D[{(\Delta\bv_\bot)}^2] = & \frac{4 \sqrt{2} \pi G^2 \rhop \mpp
                              \log\Lambda}{\sigma}
                              \bigg[\frac{\erf(X) - \mathbb{G}(X)}{X}\bigg],
\end{align}
where $X \equiv v/\sqrt{2}\sigma$ and
\begin{equation}
\label{eq:gdef}
\mathbb{G}(X)\equiv \frac{1}{2X^2}\left[\erf(X)-\frac{2X}{\sqrt{\pi}}\re^{-X^2}\right].
\end{equation}
The Cartesian diffusion coefficients are then
\begin{align}
  \label{eq:Di_c}
  D_i = {} & \frac{v_i}{v} D[\Delta v_\parallel], \\
  \label{eq:Dij_c}
  D_{ij} = {} & \frac{v_i v_j}{v^2}\big\{D[{(\Delta v_\parallel)}^2]-
                \textstyle{\frac{1}{2}}  D[{(\Delta \bv_\bot)}^2]\big\}
           +  \textstyle{\frac{1}{2}}\delta_{ij}D[{(\Delta \bv_\bot)}^2] .
\end{align}

Until now, we have considered a classical system composed of point-like
particles. We now generalize this to a system of extended field particles,
where each particle has a density profile $\rho_n(r) = \mpp W_\veps(r)$ with a
scale length $\veps$ and ${ \!\int\! \rd \br\, W_{\veps} (|\br|) = 1 }$. The
density fluctuations are now given by
$\rho(\br, t) = \mpp \sum_n W_\veps(|\br -\br_n - \bv_n t|) - \rhop$, and their
correlation function is
\begin{equation}
  \label{eq:R_soft}
  \Crho(\br, t) = \mpp\dint \bv \! \dint \brp \,  W_\veps(\br - \brp -  \bv t)W_\veps(\brp) \Fp(\bv).
\end{equation}

This approach is equivalent to using a softened version of Poisson's equation,
\begin{equation}
  \label{eq:Phi_rho_soft}
  \Phi_{\veps} (\br ,t) = -G \dint \brp \dint \brpp 
  \frac{\rho(\brpp , t)}{|\br - \brp|} \, W_{\veps} (|\brp - \brpp|),
\end{equation}
where $W_{\veps}(r)$ is the softening kernel. As discussed in
Section~\ref{sec:coulomb}, this softening cures the divergence in the Coulomb
logarithm at small scales (large wavenumbers). The Fourier transform of the
softened potential is
$\hPhi_{\veps} (\bk, \omega) = -4\pi G \, \hrho (\bk, \omega) \hW_{\veps} (k)\,
/k^{2}$. The diffusion coefficient is the same as in \eq~\eqref{eq:Dvv_2b}, but
now the Coulomb logarithm is
\begin{equation}
  \label{eq:logL_soft}
  \log\Lambda_\mathrm{soft} = \! \int_{\kmin}^{\kmax}  \frac{\rd k}{k} {|\hW_{\veps} (k)|}^2.
\end{equation}

If we take the density kernel to be Gaussian, 
\begin{equation}
  W_{\veps}(r)=\frac{1}{{(2\pi
      \veps^{2})}^{3/2}}\re^{-\frac{1}{2}r^{2}/\veps^{2}}, 
  \quad { \hW_{\veps} (\bk) = \re^{-\frac{1}{2} k^2 \veps^{2}} , }
\end{equation}
then we can let ${ \kmax \to \infty }$ and obtain
\begin{equation}
  \label{eq:logL_soft_g}
  \clog\Lambda_\mathrm{soft} =  
  \textstyle{\frac{1}{2}} \Gamma (0, \kmin^{2} \veps^{2} ),
\end{equation}
where ${\Gamma (n, x) = \int_x^\infty t^{n-1} \re^{-t} \rd t}$ is the ``upper'' incomplete Gamma function.
In the limit ${ \kmin \veps \to 0 }$, this expression is equivalent to
$\log\Lambda_\mathrm{soft}=- \log(\kmin\veps)-\frac{1}{2}\gamma_E
+\mbox{O}{(\kmin\veps)}^{2} = \log(\bmax/\veps)+\mbox{O}(1)$ with $\gamma_E$
the Euler constant. This result is consistent with the second of
\eqs~\eqref{eq:coulomb}.

For a Gaussian density kernel and a Maxwellian \DF, the density correlation
function (eq.~\ref{eq:R_soft}) is
\begin{equation}
  \label{eq:R_soft_m}
  \Crho(\br, t) = \frac{\mpp \rhob}{8\pi^{3/2} \veps^3{\Big[1 +
        {(\sigma t / \sqrt{2} \veps)}^2\Big]}^{3/2}}
\exp\Bigg[-\frac{{(r/2\veps)}^2}{1 + {(\sigma t / \sqrt{2} \veps )}^2}\Bigg].
\end{equation}
The force correlation function $\langle \bF(\br, t)\cdot \bF(\brp, \tp) \rangle = C_F(\br - \brp, t-\tp)$ is given by
\begin{equation}
  \label{eq:C_F}
  C_F(\br, t) = 4\pi G^2 \dint \brp
  \frac{\Crho(\br',t)}{|\br - \brp|} = \frac{4 \pi G^2 \mpp \rhob}{r} \erf \Bigg[\frac{r/2\veps}{\sqrt{1 + {(\sigma t / \sqrt{2} \veps )}^2}}\Bigg],
\end{equation}
which in the limit $\veps \to 0$ asymptotes to  $4 \pi G^2 \mpp \rhob r^{-1} \erf
\! \big[ r / (\sqrt{2} \sigma t) \big]$~\cite[][eq.~15]{Cohen1975}.

\subsection{Relaxation by fuzzy dark matter}
\label{sec:dc_fdm}

In this section, we describe how the stochastic density fluctuations that arise
inevitably in an \FDM\ halo lead to the diffusion of the velocity of a zero-mass
test particle.

The wavefunction ${ \psi(\br, t) }$ of the \FDM\ is governed by the
Schr\"{o}dinger--Poisson system~\citep{Ruffini+1969}
\begin{align}
  \label{eq:se}
  \ri \hbar \dpd{}{t} \psi(\br, t) 
  & 
    = -\frac{\hbar^2}{2\mb}\nabla^2 \psi(\br, t) + \mb \Phi(\br, t) \psi(\br,
    t), \\
  \label{eq:pe}
  \nabla^2 \Phi(\br, t) 
  &
    = 4\pi G {|\psi(\br, t)|}^2.
\end{align}
Here, $\mb$ is the mass of the \FDM\ particle, ${ \Phi(\br,t) }$ is the
gravitational potential, and we have assumed that the wavefunction is
normalized such that $|\psi(\br,t)|^2$ is the mass density.

To parallel our discussion of relaxation in a system of classical particles in
the preceding subsection, we assume that the self-gravity of the \FDM\ can be
ignored when considering the interaction of \FDM\ with a classical test
particle. This assumption is similar to the Jeans swindle and is valid when the
typical de~Broglie wavelength is much smaller than the scale of the system. In
this case, the \FDM\ wavefunction can be expanded as a collection of plane
waves,
\begin{equation}
  \label{eq:psi}
  \psi(\br, t) = \dint \bk \, \varphi(\bk) \, \re^{\ri \bk\cdot\br - \ri
    \omega(k) t},
\end{equation}
where
\begin{equation}
  \label{eq:omega}
  \omega(k) = \frac{\hbar k^2}{2\mb}.
\end{equation}

We assume that the ensemble averages of ${ \varphi(\bk) }$ satisfy 
\begin{equation}
  \label{eq:A_corrI}
  \langle \varphi(\bk) \rangle = 0, \;\;\; 
  \langle \varphi(\bk) \, \varphi(\bkp) \rangle
  = 0 \quad\mbox{for}\quad \bk\not=\bkp.
\end{equation}
These equations are satisfied if each plane wave has a random phase, as is
expected if they arrive in the vicinity of the test particle from large
distances and different directions. We also assume that
\begin{equation}
  \label{eq:A_corrII}
  \langle \varphi(\bk) \, \varphi^{*} (\bkp) \rangle = f_k(\bk) \, \delta(\bk - \bkp).
\end{equation}
where ${ f_k (\bk) }$ is a \DF\ defined such that the mean or ensemble-average
mass density in the volume ${ \rd \bk }$ around $\bk$ is $f_k(\bk)\rd\bk$.

These assumptions are only valid when the typical de~Broglie angular wavelength
$\lbarsig$ is much larger than the typical distance between \FDM\ particles,
$d = {(\mb/\rhob)}^{1/3}$. Therefore, the results in the remainder of this
section do not reduce to the classical diffusion coefficients in the classical
limit where $\hbar \to 0$. In the Appendix~\ref{sec:classical_lim}, we generalize
our derivation to include the classical limit. There, we show that
$\lbarsig > d$ when the \FDM\ particle mass exceeds
$m_\rs \approx {(\rhob \hbar^3/\sigma^3)}^{1/4}$, a few tens of $\eV$ in a
typical galaxy. For $\mb \gg m_\rs$ the diffusion coefficients become the
classical ones, although the system itself is not yet in the classical
limit. Classical behavior requires that the position uncertainty after a
dynamical time $T_\rd$ be small compared to the distance between particles or
$\mb \gg m_\rc$, where $m_\rc\approx {(\hbar T_\rd/2)}^{3/5}\rhob^{2/5}$, about
$10^{16} \eV$ in a typical galaxy. For any reasonable dark-matter particle
mass, the ``classical'' contribution to the relaxation is of no importance.

Note that although $\hbar$ is present in the wave function and the dispersion
relation (eqs.~\ref{eq:psi} and~\ref{eq:omega}) and thus will appear in many of
the following formulas, the following derivations can be understood entirely as
a classical field theory: the only trace of the quantum nature of the waves is
in the quadratic dispersion relation (see eq.~\ref{eq:omega}), which is rare in
classical systems.

The density fluctuations of the \FDM\ field are given by
${\rho(\br, t) = {|\psi(\br,t)|}^2 - \rhob }$, where the mean \FDM\ density is
${ \rhob = \langle {|\psi(\br,t)|}^2 \rangle = \!\!\int\!\rd \bk \, f_k(\bk)
}$. From \eq~\eqref{eq:psi}, we obtain
\begin{equation}
  \label{eq:rho_fdm}
  \rho(\br, t) = \dint \bk \dint \bkp \varphi(\bk) \, 
  \varphi^*(\bkp) \, 
  \re^{\ri (\bk - \bkp)\cdot\br - \ri [\omega(k)-\omega(\kp)]t} - \rhob.
\end{equation}

The density correlation function and its Fourier transform are
\begin{align}
  \label{eq:R_fdm}
  \Crho(\br, t) = {} & \dint \bk \dint \bkp f_k(\bk) f_k(\bkp) \,
  \re^{\ri(\bk-\bkp)\cdot\br - \ri [\omega(k)-\omega(\kp)]t}, \\
  \label{eq:Rhat_fdm}
  \hCrho (\bk, \omega) =  {} &   {(2\pi)}^4
    \dint \bkp \dint \bkpp f_k(\bk) f_k(\bkpp)
    \delta (\bk - \bkp + \bkpp) \, 
    \delta \!\left[\omega - \omega(\kp) + \omega(\kpp)\right].
\end{align}
Here, we used \eqs~\eqref{eq:A_corrI} and~\eqref{eq:A_corrII} to obtain
\begin{align}
  \label{eq:A_4pt}
  \langle \varphi(\bk_1)\varphi^*(\bk_2)\varphi^*(\bk_3)\varphi(\bk_4) \rangle
  = {} & 
         f_k(\bk_1)f_k(\bk_3)\delta(\bk_1-\bk_2)\delta(\bk_3-\bk_4)
         \nonumber \\ 
       & + f_k(\bk_1)f_k(\bk_2)\delta(\bk_1-\bk_3)\delta(\bk_2-\bk_4).
\end{align}

Each plane wave travels with velocity ${ \bv = \hbar \bk /\mb }$, and its
velocity \DF\ is given by ${ \Fb (\bv) \rd \bv = f_k (\bk) \rd \bk }$. As a
result, \eq~\eqref{eq:Rhat_fdm} can be written as
\begin{equation}
  \label{eq:Rhat_fdm_v}
  \hCrho (\bk, \omega) = 
    {(2\pi)}^4
    \dint \bv_1 \dint \bv_2 \,\Fb (\bv_1) \Fb (\bv_2)
    \delta\bigg( \bk - \frac{2\mb}{\hbar} \bvd \bigg)\,
    \delta \bigg(\omega - \frac{2 \mb}{\hbar} \bvc \ccdot \bvd \bigg),
\end{equation}
in which we have introduced the velocities
${ \bvc = (\bv_1 + \bv_2)/2}$ and
${ \bvd = (\bv_1 - \bv_2)/2}$. Note that in this case the spatial
correlation is associated with the velocity difference, while in the classical
case it is associated with the distance that a field particle travels over time
$t$ (eq.~\ref{eq:Bft}). Therefore, truncating the integrals at large and small
scales is equivalent to truncating the velocity difference.

Using \eqs~\eqref{eq:Dvv_rho} and~\eqref{eq:Rhat_fdm_v}, we obtain the diffusion
coefficient for the test particle
\begin{align}
  \label{eq:Dvv_fdmII}
  D_{ij} = {}
  &
    \frac{32\pi^3G^2\hbar^3}{\mb^3}
    \dint \bvd \dint \bvc 
    \frac{\veld^{i} \veld^{j}}{\veld^5}
    \Fb (\bvc \!+\! \bvd) \Fb (\bvc \!-\! \bvd) \,
    \delta \left( \hbvd \ccdot \bv - \hbvd \ccdot \bvc \right)
    \nonumber \\
  = {}
  &
    \frac{16\pi^3G^2\hbar^3}{\mb^3} \frac{\partial^2}{\partial v_i \partial v_j} 
    \dint \bvd \dint \bvc \, \Fb (\bvc \!+\! \bvd) \Fb (\bvc \!-\! \bvd)
    \frac{|\hbvd \ccdot \bv -  \hbvd \ccdot \bvc |}{\veld^3},
\end{align}
where $\hbvd$ is the unit vector in the direction of $\bvd$. The diffusion
coefficient $D_i$ can be obtained from $D_{ij}$ using the
fluctuation-dissipation relation (eq.~\ref{eq:Dv_fd}).

The integrals over $\bvd$ diverge logarithmically as ${ |\bvd|\to 0
}$. Therefore, we cut off the integration when ${ |\bvd|<\vdmin}$. This cutoff
arises naturally from the wave nature of \FDM\@: because $v=\hbar k/\mb$ and we
ignore wavenumbers smaller than $\kmin=1/\bmax$, we should also ignore
velocities $v \lesssim \hbar/(\mb\bmax)$. More precisely, we set
$\vdmin =\hbar/(2\mb \bmax)=\sigma \lbarsig/(2\bmax)$, where
${ \lbarsig = \hbar/(\mb\sigma) }$ is the typical de~Broglie angular
wavelength. Furthermore, the integral is dominated by the region
${ |\bvd|\ll |\bvc| }$, so we may also cut off the integration when
${ |\bvd|>v_{d,\max}}$, where ${ v_{d,\max}\lesssim \sigma }$, and approximate
$\Fb(\bvc \pm \bvd)$ by $\Fb(\bvc)$.

If we write $\bvd\equiv \veld \hbk$, the first of \eqs~\eqref{eq:Dvv_fdmII} simplifies to
\begin{align}
  \label{eq:Dvv_logL}
  D_{ij} &= {}\frac{32\pi^3G^2\hbar^3}{\mb^3} \log\Lambda_\mathrm{FDM}
    \dint \hbk\, \hk_{i} \hk_{j}\dint \bvp
    \Fb^2(\bv')
    \delta [\hbk \ccdot (\bv - \bvp) ] \nonumber \\
  & =4 G^2 \meff\,\log\Lambda_\mathrm{FDM} \dint \hbk\,\hk_{i} \hk_{j}\dint \bvp  
    \delta [\hbk \ccdot (\bv - \bvp) ]\oFb(\bvp) ,
\end{align}
and
\begin{align}
  \label{eq:Dv_logL}
  D_{i} &= 4\pi G^2 \meff \log\Lambda_\mathrm{FDM}\dpd{}{v_i} \dint \bvp \frac{\oFb(\bvp)}{|\bv - \bvp|}.
\end{align}
Here,
$\log\Lambda_\mathrm{FDM} =
\log(v_{d,\max}/\vdmin)=\log(2\bmax/\lbarsig)+\mbox{O}(1)$,
consistent with the third of \eqs~\eqref{eq:coulomb}. We have also defined
a new, effective \DF,
\begin{equation}
\oFb(\bv)=\frac{\dint \bv\,\Fb(\bv)}{\dint \bv\,\Fb^2(\bv)}\Fb^2(\bv),
\end{equation}
normalized such that $\ \dint \bv\,\oFb(\bv)=\rhob$, and an effective mass,
\begin{equation}
 \label{eq:meffI}
 \meff = \frac{{(2\pi\hbar)}^3\,\dint \bv\,\Fb^2(\bv)}{\mb^3\,\,\dint \bv\,\Fb(\bv)}.
\end{equation}
The diffusion coefficients in \eqs~\eqref{eq:Dv_logL} and~\eqref{eq:Dvv_logL}
are identical to the diffusion coefficients in \eqs~\eqref{eq:Dv_2b}
and~\eqref{eq:Dvv_2b} for classical particles, except that the particle mass
$\mpp$ is replaced by the effective mass $\meff$, the velocity \DF\ $\Fb(\bv)$
is replaced by the effective \DF\ $\oFb(\bv)$, and the Coulomb logarithm
$\log\Lambda$ is modified to $\log\Lambda_\mathrm{FDM}$. In effect, the halo
acts as if it were composed of quasiparticles with a mass $\meff$ that depends
on the local halo density and velocity distribution. These results provide a
simple recipe for computing the diffusion coefficients for a zero-mass test
particle in an \FDM\ halo.

For the Maxwellian velocity distribution (eq.~\ref{eq:maxwell}), the
integrations in \eqs~\eqref{eq:Dvv_fdmII}--\eqref{eq:meffI} can be carried out
explicitly. The effective \DF\ is a Maxwellian with the same density and a
velocity dispersion $\seff=\sigma/\sqrt{2}$. The effective mass
is
\begin{equation}
 \label{eq:meff}
  \meff = \frac{\pi^{3/2} \hbar^3 \rhob}{\mb^3 \sigma^3} =
  \rhob\, {\big(f\lsig\big)}^3,
\end{equation}
where ${ \lsig = h/(\mb\sigma) }$ is the typical de~Broglie wavelength and
$f=1/(2\sqrt{\pi})=0.282$. Moreover, by evaluating the integral in
\eq~\eqref{eq:Dvv_fdmII} for a Maxwellian, we can sharpen our estimate of the
Coulomb logarithm. We find (cf.~eq.~\ref{eq:logL_soft_g})
\begin{align}
\label{eq:logLfdm}
  \log\Lambda_\mathrm{FDM} &=\int_{\vdmin}^\infty \frac{\rd v}{v} \re^{-v^2/\sigma^2} ={\textstyle \frac{1}{2}}\Gamma(0,\vdmin^2/\sigma^2) \nonumber \\
                           &=\log(\sigma / \vdmin)  - {\textstyle \frac{1}{2}}\gamma_E + \mbox{O}{(\vdmin/\sigma)}^2.
\end{align}
Substituting $\vdmin \simeq\sigma \lbarsig /(2\bmax)$ (see the paragraph
preceding eq.~\ref{eq:Dvv_logL}),
\begin{equation}
\label{eq:logLb}
\log\Lambda_\mathrm{FDM} = {\textstyle \frac{1}{2}}\Gamma(0,{\textstyle\frac{1}{4}}\lbarsig^2/\bmax^2)
= \log(2\bmax/\lbarsig) - {\textstyle\frac{1}{2}}\gamma_E+\mbox{O}{(\lbarsig/\bmax)}^{2},
\end{equation}
which is equivalent to the classical case with a softening scale
$\veps = \lbarsig/2$ (cf.~eq.~\ref{eq:logL_soft_g}).

As the density correlation function determines the diffusion coefficients
(cf.~eqs.~\ref{eq:Dv_rho} and~\ref{eq:Dvv_rho}), it is instructive to compare
the \FDM\ correlation function to a classical one. For a Maxwellian velocity
distribution, the density correlation function (eq.~\ref{eq:R_fdm}) is given by
\begin{equation}
  \label{eq:R_fdm_m}
  \Crho(\br, t) = \frac{\rhob^2}{{\big[1 + {(\sigma t / \lbarsig)}^2\big]}^{3/2}} 
\exp\bigg[-\frac{{(r / \lbarsig)}^2}{1 + {(\sigma t / \lbarsig)}^2}\bigg].
\end{equation}
This result can be compared with numerical simulations of \FDM\
halos~\citep[e.g.,][]{Lin+2018}. Comparing this result with \eq~\eqref{eq:R_soft_m}, we see that the
density correlation function is the same as that of a classical system having a
Maxwellian \DF\ with velocity dispersion $\sigma_\rp = \seff$, a Gaussian
density kernel with a softening $\veps = \lbarsig/2$, and individual particle
mass $\mpp = \meff$. Note that uncertainties about the Coulomb logarithm are
absent from equation (\ref{eq:R_fdm_m}).

These results verify the qualitative picture of relaxation in \FDM\ halos
presented by~\citet{Hui+2017}, who assumed that the diffusion coefficients were
the same as those of a halo of classical particles with the same velocity
dispersion and effective mass $\rhob{(f_\mathrm{H}\lsig)}^3$, with
$f_\mathrm{H}\simeq 0.5$. The calculations in this section show that the actual
value of $f_\mathrm{H}$ is between 0.224 and 0.399 depending on the velocity of
the test particle and on which of the diffusion coefficient is being evaluated; thus,
the diffusion coefficients and relaxation rates are between 2 and 11 times
smaller than those assumed by~\citet{Hui+2017}. The formulation here, which defines
the effective mass for a distribution with dispersion
$\seff=\sigma/\sqrt{2}$, is both simpler and more accurate.

To estimate $\meff$, one can assume that the density is related to the radius
$r$ and the one-dimensional velocity dispersion $\sigma$ or the circular speed
$\velc$ as in a singular isothermal sphere,
\begin{equation}
\label{eq:sis}
\rhob(r)=\frac{\sigma^2}{2\pi Gr^2}=\frac{\velc^2}{4\pi Gr^2},
\end{equation}
which leads to
\begin{align}
  \label{eq:meff_approx}
  \meff = 
    \frac{\pi^{1/2} \hbar^3}{2^{1/2} G \mb^3\velc r^2}
  =     1.03\times 10^7 M_\odot \,
    \power{\frac{r}{1 \kpc}}{-2}
    \power{\frac{\mb}{10^{-22} \eV}}{-3}
    \power{\frac{\velc}{200\kms}}{-1},
\end{align}
and the typical de~Broglie wavelength is 
\begin{align}
  \label{eq:lsig}
  \lsig = \frac{h}{\mb \sigma} =
   0.85 \kpc
   \power{\frac{\mb}{10^{-22} \eV}}{-1}
    \!
    \power{\frac{\velc}{200  \kms}}{-1}.
\end{align}

\section{Dynamical friction}
\label{sec:dyf}

In the previous section, we calculated the stochastic velocity changes of a
massless particle moving through a homogeneous \FDM\ background. In this
section, we consider the additional contribution to the velocity change for a
particle of non-zero mass.

A massive particle moving through the \FDM\ field creates a gravitational wake
behind it that induces a frictional force proportional to the mass of the
particle. We will call this force dynamical friction; it is distinct from the
velocity drift described by the diffusion coefficient $D_i$ or
$D[\Delta v_\parallel]$ (eq.~\ref{eq:Dv_logL}), which is independent of the
test star's mass.\footnote{In the literature, it is common to define dynamical
  friction as the sum of both drift terms.}

The frictional force on a point object of mass $m_\rt$ traveling at velocity
$\bv_{\rt}$ through a plane wave with velocity $\bv=\hbar\bk/\mb$
is~\citep[][see also~\cite{Lora+2012}]{Hui+2017}
\begin{equation}
  \label{eq:f_fric}
  \mathbf{F_\rf} = -4\pi G^2 m^2_\rt\rho_\rb
  \frac{\bv_\rt-\bv}{{|\bv_\rt-\bv|}^3}
  \, C\left(\beta,\gamma\right),
\end{equation}
where $C(\beta, \gamma)$ is defined in~\citet[][\eq~D7]{Hui+2017} and
\begin{align}
  \label{eq:beta}
  \beta = \frac{ G\mb m_\rt}{\hbar |\bv_\rt-\bv|}, \quad 
  \gamma = \frac{\mb\bmax}{\hbar}|\bv_\rt-\bv|.
\end{align}
Here, $\bmax$ is some large radius around $m_\rt$, beyond which we assume that
the gravitational force from the medium can be ignored. Integrating
\eq~\eqref{eq:f_fric} over the \DF\ ${ \Fb (\bv) }$, we obtain the rate of
velocity drift due to dynamical friction,
\begin{equation}
  \label{eq:Dv_fric}
  D_\rf[\Delta v_i] = {}
  - 4\pi G^2 m_\rt
  \dint \bvp 
  \frac{v_{\rt,i} - \vp_i}{{|\bv_\rt-\bvp|}^3} \, \Fb (\bvp)\,
  C\left(
    \frac{G \mb m_\rt}{\hbar|\bv_\rt-\bvp|} , \frac{\mb\bmax}{\hbar}|\bv_\rt-\bvp|\right).
\end{equation}

To make an approximate estimate of the size of the quantities $\beta$ and
$\gamma$, we replace $|\bv_\rt-\bvp|$ by the velocity dispersion
$\sigma$. Then,
\begin{equation}
\gamma\approx \frac{\mb\bmax}{\hbar}\sigma= \frac{\bmax}{\lbarsig}={\textstyle\frac{1}{2}}\Lambda_\mathrm{FDM}
\end{equation}
where as usual $\lbarsig=\hbar/(\mb\sigma)$ is the typical de~Broglie angular
wavelength, and $\Lambda_\mathrm{FDM}$ is the Coulomb factor defined in
\eq~\eqref{eq:coulomb}. Similarly,
\begin{equation}
\beta\simeq \frac{G\mb m_\rt}{\hbar\sigma}=\frac{b_{90}}{\lbarsig},
\end{equation}
where $b_{90}=Gm_\rt/\sigma^2$ is the $90^\circ$ deflection radius. In the case
$\beta\gg1$, the de~Broglie wavelength is negligible, and we recover the
classical formula for dynamical friction (see~\citealt{Hui+2017}).

When $\beta\ll1$ we use the result~\citep{Hui+2017}
\begin{equation}
C(\beta,\gamma)=\mathbb{W}(\gamma)+\mbox{O}(\beta),
\end{equation}
where
\begin{equation}
\mathbb{W}(x)\equiv \mbox{Cin}(2x)+\frac{\sin(2x)}{2x}-1,
\end{equation}
and $\mbox{Cin}(x)\equiv\int_0^x (1-\cos t)\rd t/t$ is a cosine integral. Our
approximation of a homogeneous medium is only valid if the de~Broglie
wavelength is small compared to the system size, so $\gamma\gg1$ and we can use
the asymptotic expansion
$\mathbb{W}(x)\to \log(2 x) +\gamma_E-1 +\mbox{O}(1/x)$. Thus,
$C(\beta,\gamma)\simeq \log\Lambda_\mathrm{FDM}+[\log(|\bv_\rt-\bvp|/\sigma) +
\gamma_E -1]$. We may drop the term in square brackets, which is of order unity
and hence small compared to the Coulomb logarithm, and write
\begin{align}
  \label{eq:Dv_fric_a}
  D_\rf[\Delta v_i] \simeq {}
  &
    - 4\pi G^2 m_\rt \clog\Lambda_\mathrm{FDM}
    \dint \bvp \frac{v_i - \vp_i}{{|\bv - \bvp|}^3} \Fb (\bvp)
    \nonumber \\
  = {}
  &
    4\pi G^2 m_\rt \clog\Lambda_\mathrm{FDM}\dpd{}{v_i}
    \dint \bvp  \frac{1}{|\bv-\bvp|}\Fb (\bvp).
\end{align}
\Eq~\eqref{eq:Dv_fric_a} is identical to the classical formula for dynamical
friction~\citep[e.g.,][]{Tremaine+1984,Binney+2008} except that the Coulomb
logarithm is defined by the ratio of the size of the system to the de~Broglie
wavelength, rather than to the $90^\circ$ deflection radius (i.e.,
$\Lambda_\mathrm{cl}$ in eq.~\ref{eq:coulomb} is replaced by
$\Lambda_\mathrm{FDM}$). Moreover $D_\rf[\Delta v_i]$ is identical to the drift
coefficient for a test particle $D_i=D[\Delta v_i]$ (eq.~\ref{eq:Dv_2b}),
except that the particle mass $\mpp$ is replaced by the massive body's mass
$m_\rt$. For the Maxwellian velocity distribution (eq.~\ref{eq:maxwell}),
$D_\rf[\Delta v_i]=(v_i/v)D_\rf[\Delta v_\parallel]$, where
\begin{equation}
 \label{eq:Ddf}
  D_\rf[\Delta v_\parallel] =  -\frac{4\pi G^2 \rhob m_\rt \log\Lambda_\mathrm{FDM}}{\sigma^2} \,\mathbb{G}(X) ,
  \end{equation}
with $\mathbb{G}(X)$ defined in \eq~\eqref{eq:gdef}. 

\section{Mass segregation} 
\label{sec:ms}

In most current \CDM\ models, the dark matter consists of elementary particles
whose mass is negligible compared to that of any astrophysical object. Even if
the \CDM\ particles are macroscopic objects, say of $1$--$30\, M_\odot$, they
are much less massive than objects such as supermassive black holes or globular
clusters. Therefore, these objects will inspiral toward the center of a \CDM\
halo due to dynamical friction~\citep[e.g.,][]{Tremaine+1975, Begelman+1980}.
The situation is quite different in an \FDM\ halo. As shown in earlier
sections, \FDM\ behaves as if it were composed of quasiparticles with an
effective mass given by \eq~\eqref{eq:meff_approx}. Thus, although the massive
object still loses orbital energy by dynamical friction, it can also can gain
energy by gravitational interactions with the quasiparticles. We expect that
the inspiral of the massive object will stall if it reaches energy
equipartition with the quasiparticles. For similar reasons, individual stars
will tend to gain energy from interactions with the \FDM\ quasiparticles, and
this process can lead to the expansion of a stellar system embedded in the
halo.
 
To explore these processes quantitatively, we use a simple model of a galaxy
containing only \FDM\@, with density $\rhob$ and a Maxwellian velocity
distribution with dispersion $\sigma$ (eq.~\ref{eq:maxwell}). Then, we can
combine \eqs~\eqref{eq:Dvv_logL}--\eqref{eq:Dv_logL} and~\eqref{eq:Ddf} to
obtain the diffusion coefficients for a point mass $m_\rt$,
\begin{align}
  \label{eq:Dpar_f}
  D[\Delta v_\parallel] = & -\frac{4\pi G^2 \rhob \meff \log\Lambda_\mathrm{FDM}}{\seff^2}
                            \big[\mathbb{G}(\Xeff)   +\mueff\, \mathbb{G}(X)], \\
  \label{eq:D2para}
  D[{(\Delta v_\parallel)}^2] = & \frac{4 \sqrt{2} \pi G^2 \rhob \meff
    \log\Lambda_\mathrm{FDM}}{\seff}
  \frac{\mathbb{G}(\Xeff)}{\Xeff}, \\
  \label{eq:D2pera}
  D[{(\Delta \bv_\bot)}^2] = & \frac{4 \sqrt{2} \pi G^2 \rhob \meff
                               \log\Lambda_\mathrm{FDM}}{\seff}
                               \frac{\erf(\Xeff) - \mathbb{G}(\Xeff)}{\Xeff},
\end{align}
where $\mathbb{G}(X)$ is defined in \eq~\eqref{eq:gdef},
$\seff^2/\sigma^2 = 1/2$, $X = v/\sqrt{2}\sigma$,
$\Xeff = v/\sqrt{2}\seff = v/\sigma$, and 
\begin{equation}
  \label{eq:mu}
  \mueff \equiv  \frac{m_\rt}{\meff}\frac{\seff^2}{\sigma^2} = \frac{m_\rt}{2\meff}
\end{equation}
is the effective mass ratio. Note the factor of 2 in the definition of
$\mueff$, and note also that the classical diffusion coefficient analogous to
\eq~\eqref{eq:Dpar_f} for a halo composed of particles of mass $\mpp$ is
\begin{equation}
\label{eq:df_cdm}
 D[\Delta v_\parallel] = -\frac{4\pi G^2 \rhop \mpp \log\Lambda_\mathrm{cl}}{\sigma^2}
                            (1+\mu_\mathrm{cl})\mathbb{G}(X),
\end{equation}
where $\mu_\mathrm{cl}=m_\rt/\mpp$ (without a factor of 2).

We stress again that the diffusion coefficients in
\eqs~\eqref{eq:Dpar_f}--\eqref{eq:D2pera} do not go to the classical ones in
the limit $\hbar \to 0$ (or $\meff \to 0$). This incompleteness is related to
our simplifying assumption about the wave function in Section~\ref{sec:dc_fdm}
(see discussion after eq.~\ref{eq:A_corrII}). In the
Appendix~\ref{sec:classical_lim} we extend the derivation of
Section~\ref{sec:dc_fdm} to include the classical limit. There, we show that
when $\mb \gg \meff$, the relaxation becomes the classical one (i.e., as in a
system of classical particles of mass $\mb$ with velocity dispersion
$\sigma$). This ``classical'' correction is negligible for the \FDM\ mass
considered here, $\mb < 10^{-20} \eV$, for which we can expect that the
dynamics will deviate from the standard \CDM\ case.

The specific energy diffusion coefficients are
\begin{align}
  \label{eq:DE_fdm_m}
  D[\Delta E] = {}
  &
    vD[\Delta v_\parallel] + \textstyle{\frac{1}{2}} D[{(\Delta v_\parallel)}^2] +
    \textstyle{\frac{1}{2}} D[{(\Delta \bv_\bot)}^2]
    \nonumber \\
  = {} 
  &
    \frac{4 \sqrt{2\pi} G^2 \rhob \meff \log\Lambda_\mathrm{FDM}}{\seff}
    \big[
    \exp(-\Xeff^2) - \mueff \sqrt{\pi} \Xeff \mathbb{G}(X)\big]
    \nonumber \\
  = {} 
  &
    \frac{8 \sqrt{\pi} G^2\rhob\meff\log\Lambda_\mathrm{FDM}}{\sigma}\,
    \exp(-v^2/\sigma^2) \big[1- \mueff K(v/\sigma)\big],
\end{align}
and
\begin{equation}
  \label{eq:DEE_fdm_m}
  D[{(\Delta E)}^2] = 
    v^2D[{(\Delta v_\parallel)}^2] =
    8 \sqrt{2} \pi G^2 \rhob \meff \seff \log\Lambda_\mathrm{FDM}\Xeff \mathbb{G}(\Xeff),
\end{equation}
where we defined the dimensionless function
\begin{equation}
  \label{eq:Kdef}
  K(x)\equiv \frac{\sqrt{\pi}}{x}\re^{x^2}\erf(x/\sqrt{2})
    - \sqrt{2}\re^{x^2/2}.
\end{equation}

The mean change in energy (eq.~\ref{eq:DE_fdm_m}) arises from the competition
between two processes: (i) diffusion (``heating''), a term resulting from the
potential fluctuations of the \FDM\ field that is proportional to $\meff$ and
pumps energy into the orbit of the massive body, and (ii) dynamical friction
(``cooling''), a term resulting from the back-reaction of the massive body on
the \FDM\ that is proportional to the body's mass $m_\rt$ and transfers energy
from its orbit into the \FDM\ field. The ratio between cooling and heating is
given by $\mueff K(v/\sigma)$.

To investigate this process in more detail, let us consider an ensemble of
systems, each containing a single body of mass $m_\rt$ traveling in a
uniform background of \FDM\@. The velocities of these bodies are distributed
according to a Maxwellian \DF\ $F_\rt(\bv_\rt)$, analogous to
\eq~\eqref{eq:maxwell} but with density and velocity dispersion $\rho_\rt$ and
$\sigma_\rt$. The flow of specific energy into the orbits of the massive
objects is
\begin{align}
  \label{eq:Edot}
  \langle \dot{E} \rangle = {} 
  &
    \frac{1}{\rho_\rt} \dint \bv_\rt F_\rt(\bv_\rt) D[\Delta E]
  \nonumber \\
 = {}
&
  \frac{8 \sqrt{\pi} G^2 \rhob \meff \log\Lambda_\mathrm{FDM}}{\sigma {(1+2\sigma_\rt^2/\sigma^2)}^{3/2}}
  \bigg[1
 - \mueff \frac{\sqrt{2}\sigma_\rt^2}{\sigma^2} 
  \frac{{(1+2\sigma_\rt^2/\sigma^2)}^{3/2}}{{(1+\sigma_\rt^2/\sigma^2)}^{3/2}}
  \bigg].
\end{align}

When $m_\rt\ll \meff$ or $\mueff \ll 1$, heating dominates, and we can use
\eq~\eqref{eq:Edot} to write
\begin{equation}
  \label{eq:sigma_t_dot}
  \od{\sigma_\rt^2}{t} = {\textstyle \frac{2}{3}} \langle \dot{E} \rangle  
  = \frac{\sigma^2}{T_\mathrm{heat}} {(1+2\sigma_\rt^2/\sigma^2)}^{-3/2},
\end{equation}
where we defined 
\begin{equation}
  \label{eq:t_heat}
  T_\mathrm{heat} =
  \frac{3\sigma^3}{16 \sqrt{\pi} G^2 \rhob \meff \log\Lambda_\mathrm{FDM}}
   \! = \!
  \frac{3\mb^3 \sigma^6}{16 \pi^2 G^2 \rhob^2 \, \hbar^3 \log\Lambda_\mathrm{FDM}},
\end{equation}
as the heating timescale.
The solution to \eq~\eqref{eq:sigma_t_dot} is 
\begin{equation}
  \label{eq:sigma_t}
  \frac{\sigma_\rt^2(t)}{\sigma^2} ={\textstyle\frac{1}{2}}
  {\left\{5t/T_\mathrm{heat} +
    {[1+2\sigma_\rt^2(0)/\sigma^2]}^{5/2} 
  \right\}}^{2/5} -\textstyle\frac{1}{2}.
\end{equation}
Therefore, at time $t$, the velocity dispersion $\sigma_\rt^2$ should be at
least $\frac{1}{2}{(5 t /T_\mathrm{heat} + 1)}^{2/5}-\frac{1}{2}$ times
$\sigma^2$, and $\sigma_\rt$ will exceed $\sigma$ in a time
$t\lesssim 3 T_\mathrm{heat}$.

When $\mueff \gg 1$, cooling dominates, and \eq~\eqref{eq:Edot} can be written
as
\begin{equation}
  \label{eq:sigma_t_dota}
  \od{\sigma_\rt^2}{t} = -\frac{\sigma_\rt^2}{T_\mathrm{cool}}{(1+\sigma_\rt^2/\sigma^2)}^{-3/2},
\end{equation}
in which we defined 
\begin{equation}
  \label{eq:t_cool}
  T_\mathrm{cool} = \frac{3 \sigma^3 }{8\sqrt{2\pi} G^2 m_\rt\,
    \rhob \log\Lambda_\mathrm{FDM}},
\end{equation}
as the cooling time. As expected, $T_\mathrm{cool}$ is independent of the
effective mass of the \FDM\ and is identical to the classical result except for
a change in the Coulomb logarithm.

When $T_\mathrm{heat}$ and $T_\mathrm{cool}$ are smaller than the lifetime of
the system, the distribution of velocities of the ensemble of massive objects
will relax to a steady state, which is determined by requiring that its \DF\
$F_\rt(\bv)$ satisfy the zero-flux condition in energy space:
\begin{equation}
  \label{eq:1}
  \od{}{v} \Big\{ D[{(\Delta E)}^2] \, v \, F_\rt (v) \Big\} = 2 v^2 D[(\Delta
  E)] F_\rt (v).
\end{equation}
This is solved to give 
\begin{equation}
  \label{eq:df_ss}
  F_\rt (v) \propto
  \frac{1}{v D[{(\Delta E)}^2]}
  \, \exp \int_0^v \rd \vp  \frac{2\vp\,D[\Delta E]}{D[{(\Delta E)}^2]} ,
\end{equation}
with the normalization chosen so that ${ \!\int\! \rd\bv\,F_\rt(v)=\rho_\rt }$.

\begin{figure}[ht]
\plotone{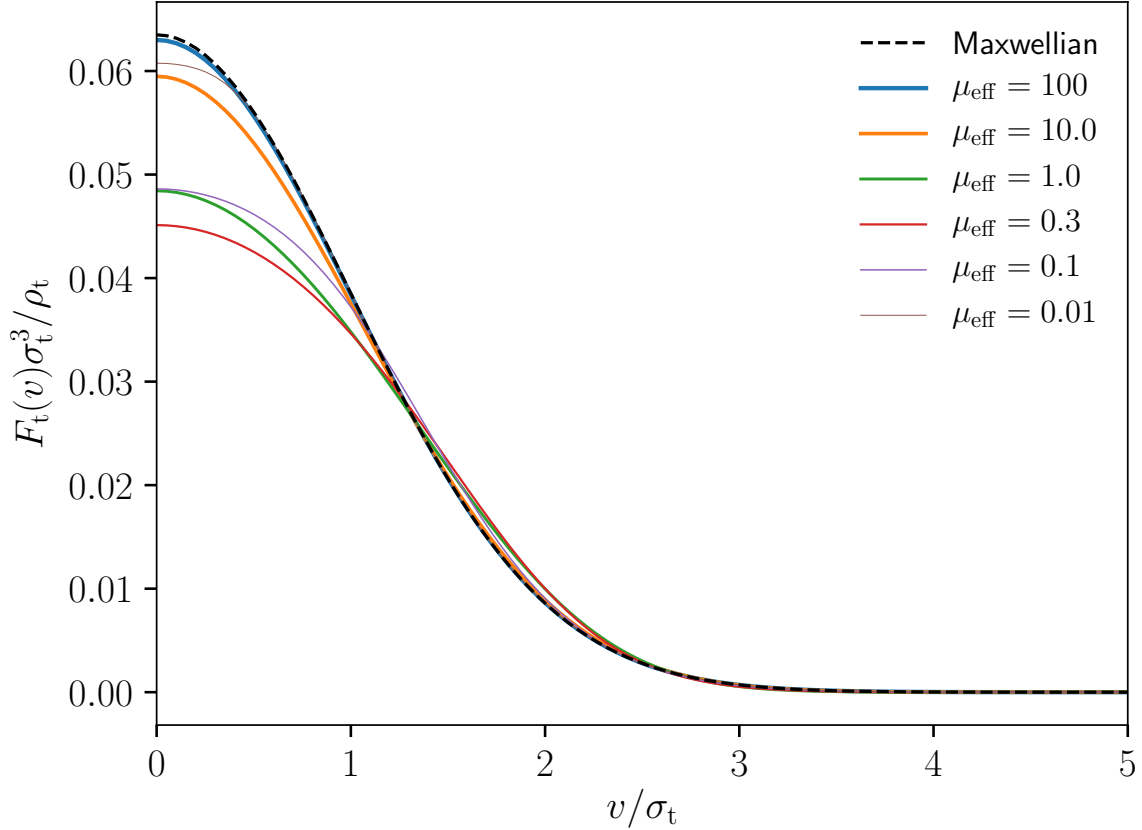}
\caption{The steady-state velocity distribution for an ensemble of massive
  bodies interacting with an \FDM\ field, as obtained from \eq~\eqref{eq:df_ss}
  for several values of the effective mass ratio
  $\mueff\equiv m_\rt/(2 \meff)$ (solid lines). The velocity is plotted in units of
  the velocity dispersion, $\sigma_\rt$, where
  $\sigma_\rt^2=\langle v^2\rangle/3$. The steady-state velocity distribution
  approaches  a Maxwellian (dashed line) in the limits $\mueff \to 0$ and
  $\mueff \to \infty$.\label{fig:dfv}}
\end{figure}

\begin{figure}[ht]
\plotone{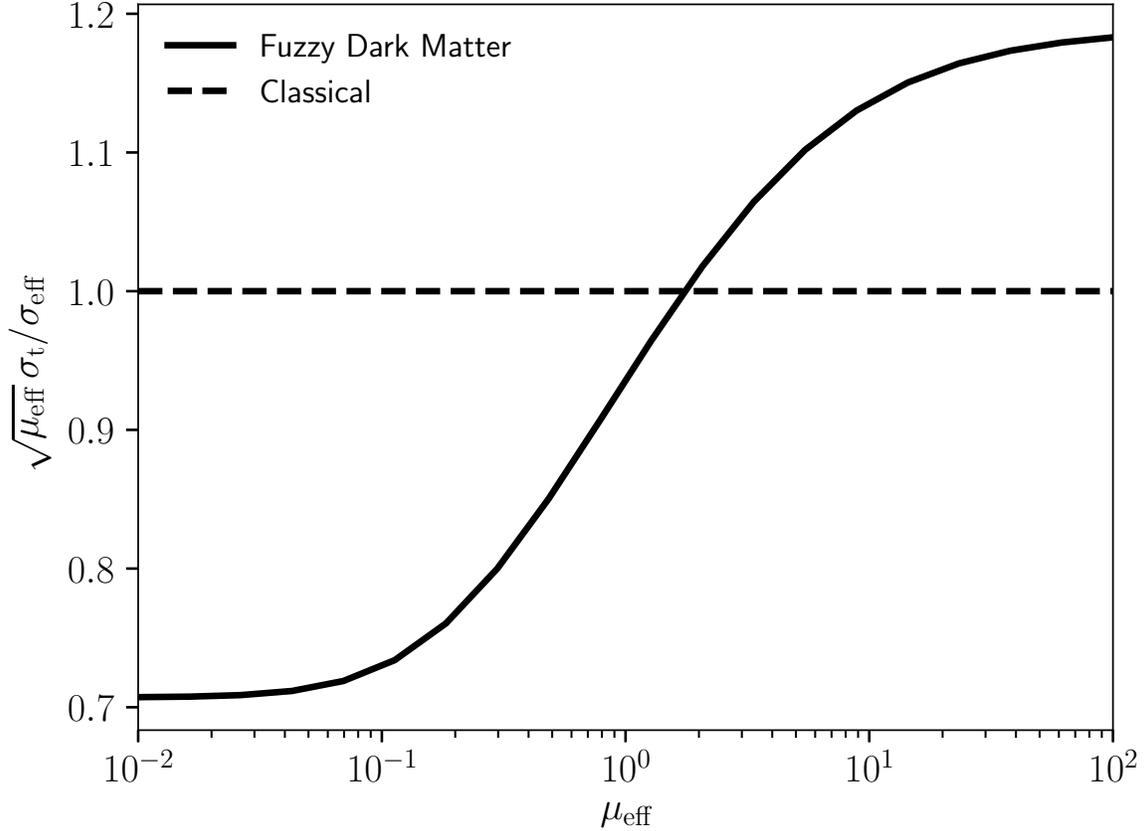}
\caption{In thermal equilibrium, the velocity dispersion $\sigma_\rt$ of an
  ensemble of baryons is related to the effective velocity dispersion
  $\seff = \sigma/\sqrt{2}$ of the \FDM\ halo. The ratio of
  dispersions depends on the effective mass ratio
  $\mueff\equiv m_\rt/(2\meff)$ (solid line) and is close to (within
  30\%) but not equal to the standard relation
  $\sqrt{m_\rt/\mpp} \, \sigma_\rt/\sigma= 1$ (dashed line) that corresponds to the
  thermal equilibrium in a background of classical particles of mass $\mpp$ and
  velocity dispersion $\sigma$.\label{fig:sigma}}
\end{figure}

As the diffusion coefficients depend linearly on the halo mass density $\rhob$,
the velocity distribution ${ F_\rt (v) }$ depends on $\rhob$ only through
$\mueff$, via the dependence of the effective mass $\meff$ on $\rhob$
(eq.~\ref{eq:meff}). For a classical $N$-body system composed of particles of
mass $\mpp$, ${ F_\rt (v) }$ is a Maxwellian with mean-square velocity
$3(\mpp/m_\rt)\sigma^2$. In contrast, the steady-state velocity distribution of
an ensemble of massive bodies interacting gravitationally with the \FDM\ field
is only approximately Maxwellian (see Figure~\ref{fig:dfv}), although the
mean-square velocity is close to $3\seff^2/\mueff=3(\meff/m_\rt)\sigma^2$,
similar to the classical relation (see Figure~\ref{fig:sigma}).

\subsection{Examples}

We now give two examples of the dynamical interaction between an \FDM\ halo the
and baryonic objects orbiting within it. These examples are based on a
simplified model of the \FDM\ halo, consisting of two components:

\paragraph{The central soliton} Near the center, the \FDM\ is condensed into a
soliton, which is the ground-state solution of the Schr\"{o}dinger--Poisson
equations. The density of the soliton can be approximated
by~\citep{Schive+2014a}
\begin{equation}
  \label{eq:rhos}
  \rho_\rs(r) \approx  \frac{0.019 M_\odot \pc^{-3}}{{[1 + 0.091{(r/r_\rs)}^2]}^8}
  \power{\frac{\mb}{10^{-22}\, \mathrm{eV}}}{-2}
  \power{\frac{r_\rs}{\kpc}}{-4}.
\end{equation}
The total soliton mass is 
\begin{equation}
  \label{eq:M_s}
  M_\rs=4\pi\int_0^\infty r^2 \rd r\,\rho_\rs(r) 
  = 2.2\times 10^8M_\odot\,\power{\frac{r_\rs}{\kpc}}{-1}
  \power{\frac{\mb}{10^{-22}\,\mathrm{eV}}}{-2}.
\end{equation}
Numerical simulations of the evolution of \FDM\ halos in a cosmological context
find that the soliton core radius $r_\rs$ is related to the total halo (virial)
mass $M_\rh$ by~\citep{Schive+2014b}
\begin{equation}
  \label{eq:core}
  r_\rs \simeq
  0.16\kpc\,{\left(\frac{\mb}{10^{-22}\eV}\right)}^{-1}{\left(\frac{M_\rh}{10^{12}M_\odot}\right)}^{-1/3},
\end{equation}
The relation between halo mass and peak circular speed $v_{\max}$ outside the
soliton is the same as in \CDM\ \citep{Klypin+2011},
\begin{equation}
  \label{eq:vmax}
  v_{\max}=155\kms\,\power{\frac{M_\rh}{10^{12}M_\odot}}{0.316},
\end{equation}
so the relation between the soliton core radius and the peak circular speed is
\begin{equation}
  \label{eq:corev}
  r_\rs \simeq
  0.12\kpc\,{\left(\frac{\mb}{10^{-22}\eV}\right)}^{-1}\power{\frac{v_{\max}}{200\kms}}{-1.05}.
\end{equation}

It is instructive to compare the typical de~Broglie wavelength in the galaxy to
the radius of the soliton, and we can do this in two ways. (i) The wavelength
for a particle traveling at the circular speed at a distance $r$ outside the
soliton is given by the simple formula
\begin{equation}
  \label{eq:solrad}
  \lambda=\frac{h}{\mb}\power{\frac{r}{GM_\rs}}{1/2}=3.91 \, r_\rs\power{\frac{r}{r_\rs}}{1/2}.
\end{equation}
In other words, the de~Broglie wavelength just outside the soliton is on the
order the soliton radius. (ii) The de~Broglie wavelength for a particle
traveling at the peak circular speed is
\begin{equation}
  \label{eq:solrad1}
  \lambda=\frac{h}{\mb v_{\max}}=0.60\kpc {\left(\frac{\mb}{10^{-22}\eV}\right)}^{-1}\power{\frac{v_{\max}}{200\kms}}{-1}.
\end{equation}
Once again, the de~Broglie wavelength is a few times the soliton radius
(eq.~\ref{eq:corev}). The agreement between methods (i) and (ii) reflects the
fact that the empirical relation~\eqref{eq:core} implies that the peak circular
speed in the soliton is almost the same (25\% smaller) as the peak circular
speed in the halo, independent of the particle mass and almost independent of
the halo mass. This coincidence is an unexplained feature of the evolution of
\FDM\ halos.

\paragraph{The halo} Outside the soliton, the mean \FDM\ density distribution is
expected to be similar to that of \CDM\ halos, which can be fit empirically by
the~\citet{nfw1997} profile. We shall adopt an even simpler model, in which
outside the soliton, the \FDM\ density is given by a singular isothermal sphere
(eq.~\ref{eq:sis}). The effective mass (eq.~\ref{eq:meff}) of the \FDM\ field
at radius $r\gg r_\rs$ is then given by \eq~\eqref{eq:meff_approx}, and the
typical de~Broglie wavelength is given by \eq~\eqref{eq:lsig}.

\subsubsection{Inspiral of a massive object}

An object of mass $m_\rt$ on a circular orbit of initial radius $\rri$ will
inspiral toward the central soliton if the effective mass ratio $\mueff \gg 1$
(eq.~\ref{eq:DE_fdm_m}). The inspiral time is~\citep[][eq.\ 8.12]{Binney+2008}
\begin{align}
  \label{eq:tf}
  t_\mathrm{inspiral} = {}
  &
  \frac{1.65 \, \rri^2 \sigma}{\log\Lambda_\mathrm{FDM} G m_\rt}
    \nonumber \\
  = {}
  &
    \frac{84.9 \Gyr}{\log \Lambda_\mathrm{FDM}}
    \frac{10^7 M_\odot}{m_\rt}
    \frac{\velc}{200  \kms}
    \power{\frac{\rri}{4 \kpc}}{2}.
\end{align}

The object will spiral to the center in less than the age of the galaxy,
$T_\mathrm{age}$, if
\begin{equation}
  \label{eq:ri}
  \rri <  2.08\kpc\,
  \power{\frac{\log\Lambda_\mathrm{FDM}}{\log 10}
  \frac{m_\rt}{10^7 M_\odot}
  \frac{T_\mathrm{age}}{10 \Gyr}
  \frac{200\kms}{\velc}}{1/2}.
\end{equation}
However, as the radius of the orbit shrinks, the effective mass grows as
$r^{-2}$ (eq.~\ref{eq:meff_approx}), so $\mueff \propto r^2$. The effective mass
ratio is less than unity inside a stalling radius
\begin{align}
  \label{eq:r1}
  r_\mathrm{stall} =  
  &
    {{(2\pi)}^{1/4}\left(\frac{\hbar^3}{Gm_\rt\mb^3\velc}\right)}^{1/2}
    =1.43\kpc\,\power{\frac{m_\rt}{10^7 M_\odot}}{-1/2}
    \power{\frac{\mb}{10^{-22} \eV}}{-3/2}
    \power{\frac{\velc}{200 \kms}}{-1/2}.
\end{align}

In early-type galaxies (ellipticals and spiral bulges), the mass of the central
black hole is correlated with the velocity dispersion~\citep{Kormendy+2013},
\begin{equation}
  \label{eq:msig}
  \log_{10} \frac{M_\bullet}{10^9M_\odot}=-0.51\pm0.05+(4.4\pm0.3)\log_{10}\frac{\sigma}{200\kms}
\end{equation}
with a scatter of about 0.3 dex. If we assume that the circular speed and
dispersion are related by $\sigma=\velc/\sqrt{2}$ as in the isothermal sphere,
and that the mass of the inspiraling black hole is $m_\rt=fM_\bullet$ with $f<1$,
then these relations can be rewritten as
\begin{equation}
  \label{eq:ri_msig}
  \rri <  3.1\kpc\,
  \power{\frac{f}{0.1}\frac{\log\Lambda_\mathrm{FDM}}{\log 10}
  \frac{T_\mathrm{age}}{10 \Gyr}}{1/2}\power{\frac{\sigma}{200\kms}}{1.7};
\end{equation}
\begin{equation}
  \label{eq:r1_msig}
  r_\mathrm{stall} =  0.70\kpc\,\power{\frac{f}{0.1}}{-1/2}\power{\frac{\mb}{10^{-22} \eV}}{-3/2}
    \power{\frac{\sigma}{200 \kms}}{-2.7}.
\end{equation}

These results suggest that the inspiral of supermassive black holes in \FDM\
halos may be stalled at orbital radii of a few hundred parsecs, a possibility
that has been discussed already by~\citet{Hui+2017}. Although we believe that
the physical mechanism described here is robust, there are two (related)
shortcomings in these calculations: (i) for the parameters of interest, the
stalling radius can be comparable to or even smaller than the typical
de~Broglie radius $\lsig$ (eq.~\ref{eq:lsig}); because the maximum scale of the
encounters for which our approximations are valid is then
$\bmax\simeq r_\mathrm{stall}$, the argument of the Coulomb logarithm is
$\Lambda_\mathrm{FDM}=2r_\mathrm{stall}/\lbarsig$ (eq.~\ref{eq:logLb}), small
enough that the assumption $\Lambda_\mathrm{FDM} \gg 1$ on which our
calculations are based is suspect; (ii) the stalling radius is not much larger
than the core radius of the central soliton $r_\rs$ (eq.~\ref{eq:core}), and
inside the soliton, heating by fluctuations in the \FDM\ vanishes\footnote{This
  conclusion assumes that the soliton is in its ground state. Simulations
  by~\citet{Veltmaat+2018} suggest that the soliton typically exhibits strong
  density oscillations, which could add energy to nearby orbits.} although
dynamical friction does not. These limitations are related because the
de~Broglie wavelength just outside the soliton is of order the soliton radius
(eqs.~\ref{eq:solrad} and~\ref{eq:solrad1}).

\begin{figure}[ht]
\plotone{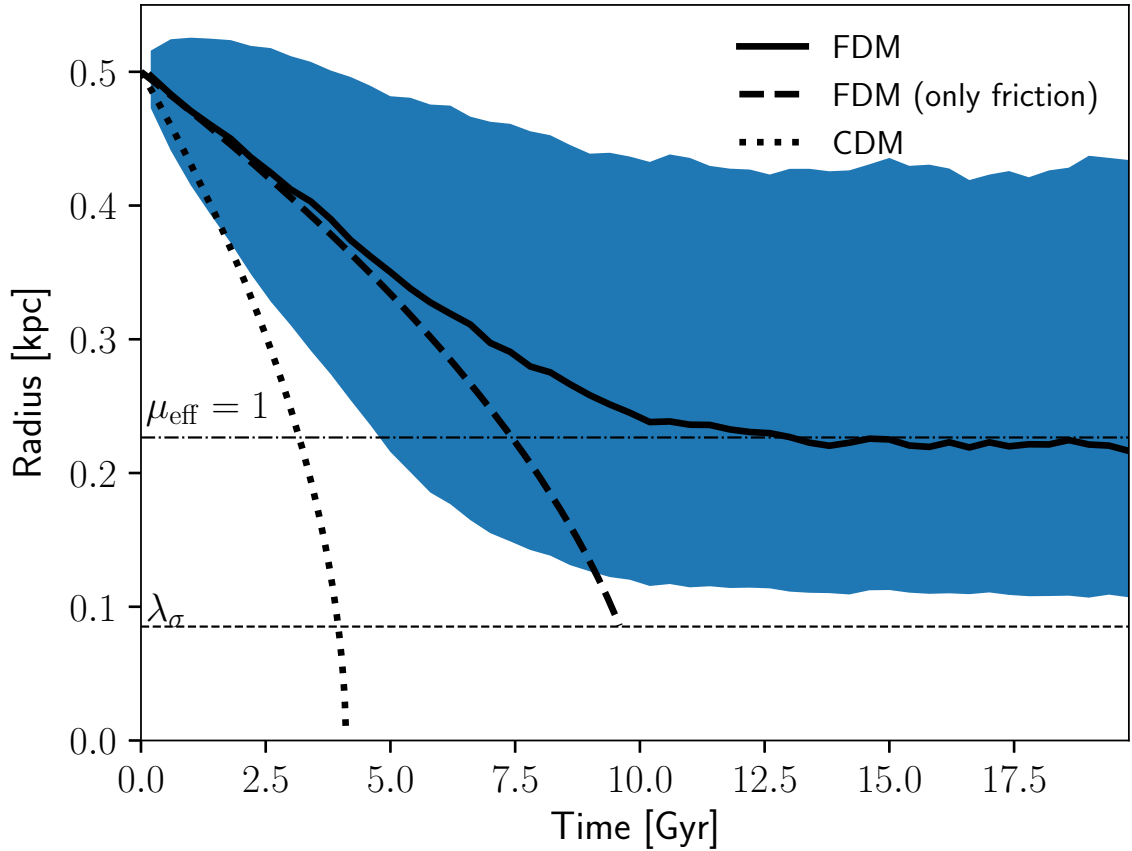}
\caption{The inspiral of a massive object ($m_\rt=4\times10^5 M_\odot$) on a
  circular orbit in a spherical galaxy with constant circular speed
  $\velc = 200 \kms$ (eq.~\ref{eq:sis}). The dotted line shows the evolution of
  the orbital radius due to dynamical friction if the galaxy is composed of
  \CDM\ (eq.~\ref{eq:df_cdm} with $\mu_\mathrm{cl}\gg1$). The dashed line shows
  the evolution due to dynamical friction if the galaxy is composed of \FDM\
  (eq.~\ref{eq:Ddf}) and diffusion terms are ignored. This differs from the
  \CDM\ case only through the Coulomb logarithm. The solid line and shaded
  region show the evolution in an \FDM\ galaxy including both dynamical
  friction and diffusion, assuming an \FDM\ mass $\mb=10^{-21}\eV$. We have
  carried out 1000 realizations of the orbital evolution, and the solid line
  and shaded region show the median and central $68\%$ region. The median
  radius saturates close to where $\mueff = 1$ (dashed-dotted horizontal
  line). This behavior is different from the case where diffusion is ignored
  (dashed line), for which dynamical friction causes the orbit to decay at
  least down to the de~Broglie wavelength $\lsig$ (dashed horizontal
  line).\label{fig:friction}}
\end{figure}

In Figure~\ref{fig:friction}, we illustrate how energy diffusion due to
scattering by \FDM\ quasiparticles tampers with the otherwise deterministic
inspiral due to dynamical friction.
We followed the orbit of a massive object ($m_\rt = 4\times 10^5M_\odot$)
in a singular isothermal sphere (eq.~\ref{eq:sis}) having circular speed
$\velc = 200\kms$, and applied random velocity changes using the diffusion
coefficients~\eqref{eq:Dpar_f}--\eqref{eq:D2pera} with $\mb = 10^{-21} \eV$. We
repeated this process $1000$ times, and Figure~\ref{fig:friction} shows the
median and 68\% confidence band of the orbital radius as a function of time.
For comparison, we also applied (deterministic) velocity changes due to
dynamical friction only (eq.~\ref{eq:Ddf}) both for \FDM\ and \CDM\ halos,
which differ only in the Coulomb logarithm. The results shown in
Figure~\ref{fig:friction} are consistent with our claim that a massive object
that is inspiraling to the center by dynamical friction will tend to stall, on
average, at a radius where the effective mass ratio $\mueff \simeq 1$.

\begin{figure}[ht]
\plotone{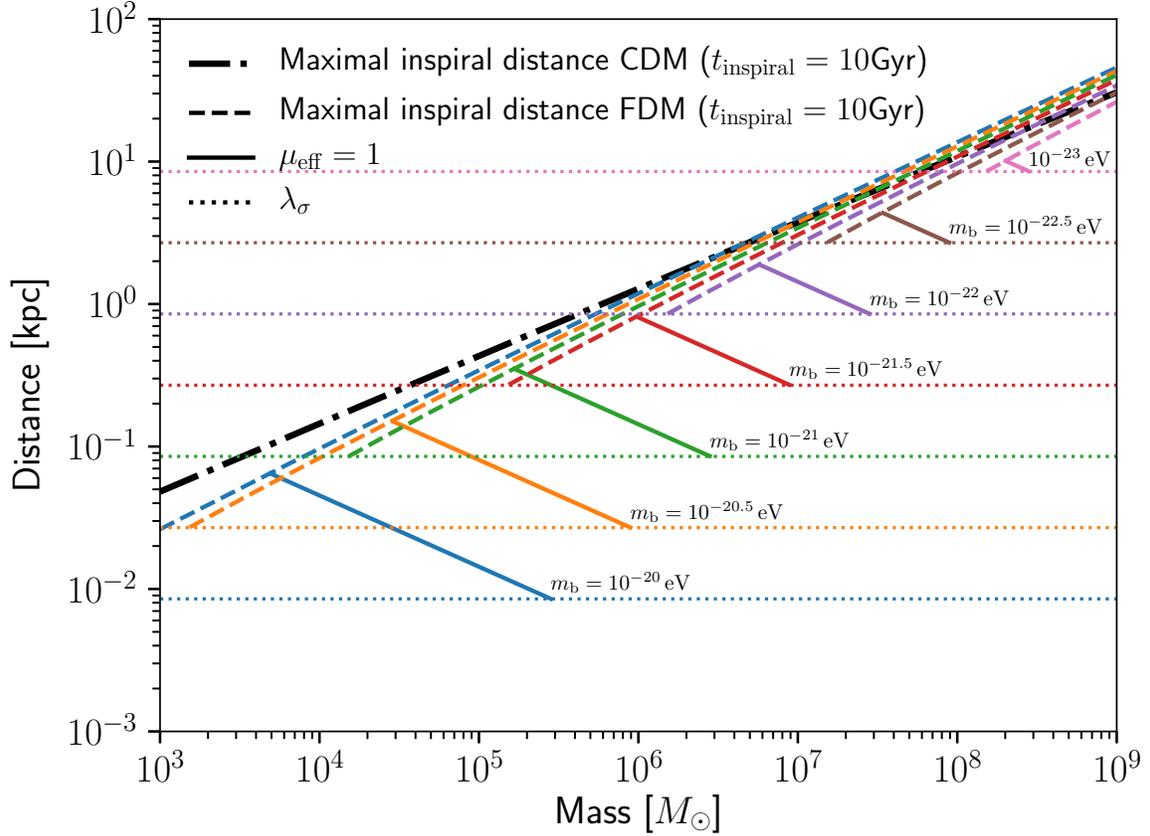}
\caption{A massive object initially on a circular orbit will spiral to the
  center within ${ 10 \Gyr }$ if it lies below the dashed lines
  (eq.~\ref{eq:ri}). Different colors represent different assumptions about the
  mass of the \FDM\ particle, and the heavy dashed-dotted black line shows the
  same curve for \CDM\@. The solid lines show the radius
  $r_\mathrm{stall}$ (eq.~\ref{eq:r1}) where stochastic potential fluctuations
  cause the inspiral to stall. The solid lines terminate when the stalling
  radius is smaller than the typical de~Broglie wavelength $\lsig$ (dotted
  lines); at smaller distances,  the evolution is dominated by interactions with
  the soliton, which exerts dynamical friction but has no potential
  fluctuations. We assume that the density of the \FDM\ is that of a singular
  isothermal sphere (eq.~\ref{eq:sis}) with circular speed
  ${ \velc = 200 \kms }$.\label{fig:r0_r1}}
\end{figure}

\begin{figure}[ht]
\plotone{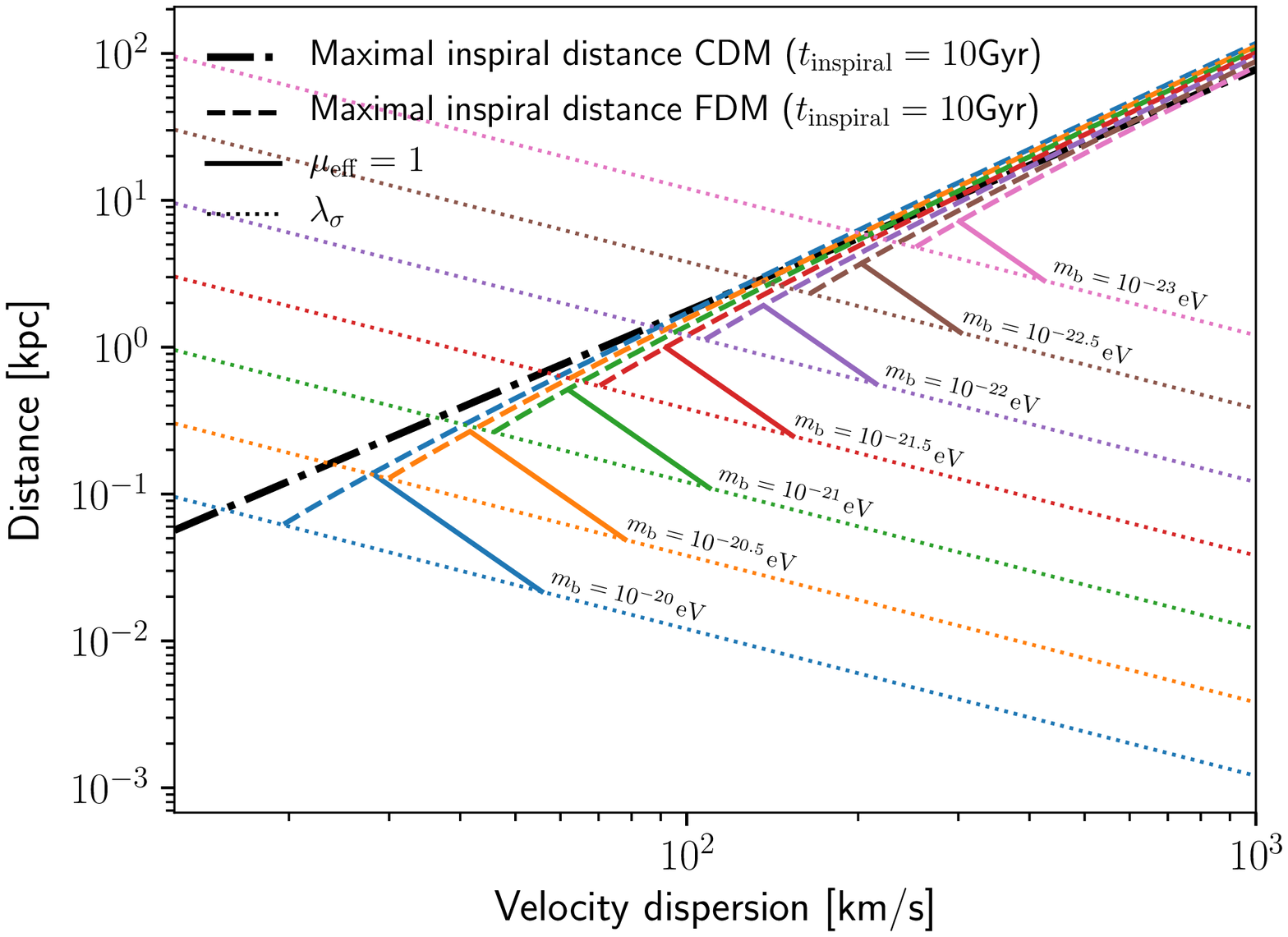}
\caption{Same as Figure~\ref{fig:r0_r1}, except that the horizontal axis is the
  velocity dispersion of the host galaxy and the mass of the inspiraling object
  is a fraction $f=0.1$ of the mass of the central black hole inferred from the
  ${M\!-\!\sigma}$ relation (eq.~\ref{eq:msig}).\label{fig:rstall_msig}}
\end{figure}

In Figure~\ref{fig:r0_r1}, we show the relation between the maximum inspiral
distance $\rri$ (eq.~\ref{eq:ri}), the stalling radius $r_\mathrm{stall}$
(eq.~\ref{eq:r1}), and the typical de~Broglie wavelength $\lsig$
(eq.~\ref{eq:lsig}) of a galaxy with circular speed $\velc = 200 \kms$, for a
range of massive objects and \FDM\ particle masses. Similarly, in
Figure~\ref{fig:rstall_msig}, we show the relation between the maximum inspiral
distance $\rri$ (eq.~\ref{eq:ri_msig}), the stalling radius $r_\mathrm{stall}$
(eq.~\ref{eq:r1_msig}), and the typical de~Broglie wavelength $\lsig$
(eq.~\ref{eq:lsig}) for a massive object that is a fraction $f=0.1$ of the
central black hole mass inferred from the $M$--$\sigma$ relation
(eq.~\ref{eq:msig}) for galaxies with a range of velocity dispersions and \FDM\
particle masses.

\subsubsection{Heating of a spherical stellar population}

We consider the effect of \FDM\ fluctuations on a stellar system having a
Maxwellian \DF\ with velocity dispersion $\sigma_\rt$. We assume that the
gravitational potential is dominated by \FDM\ and that the typical radius of
the stellar system is $r_\star$. Since $\meff$ is much larger than the mass of
any individual star, dynamical friction and cooling are negligible. The heating
timescale is given by \eqs~\eqref{eq:t_heat} and~\eqref{eq:sis}:
\begin{align}
  \label{eq:t_heat_bis}
  T_\mathrm{heat} = {}
  &
  \frac{3\mb^3 \velc^2 r_\star^4}{8 \hbar^3 \log\Lambda_\mathrm{FDM}}
    \nonumber \\
  \approx {}
  &
    \frac{2.08 \Gyr}{\log\Lambda_\mathrm{FDM}}
    \power{\frac{r_\star}{1  \kpc}}{4} 
    \power{\frac{\mb}{10^{-22}  \eV}}{3}
    \power{\frac{\velc}{200  \kms}}{2}.
\end{align}
The heating will be significant if $T_\mathrm{heat}$ is less than a third of
the age of the galaxy $T_\mathrm{age}$ (see discussion following
eq.~\ref{eq:sigma_t}), which occurs if $r_\star < r_\mathrm{heat}$ where the
heating radius is
\begin{equation}
  \label{eq:rheat}
  r_\mathrm{heat}=1.13\kpc{\left(\log\Lambda_\mathrm{FDM}
      \frac{T_\mathrm{age}}{10\Gyr}\right)}^{1/4}{\left(\frac{\velc}{200\kms}\right)}^{-1/2}
  {\left(\frac{\mb}{10^{-22}\eV}\right)}^{-3/4}.
\end{equation}
The approximations we are using are only valid if the orbital radius is
significantly larger than the de~Broglie wavelength. Setting
$r_\star=\lsig$ we obtain a minimum heating time
\begin{align}
  \label{eq:t_heat_min}
  T_\mathrm{heat}^{\min} = {}
  &
    \frac{24 \pi^4 \hbar }{\mb \velc^2\log\Lambda_\mathrm{FDM}}
    \nonumber \\
  \approx {}
  &
    0.43 \Gyr
    \power{\frac{\mb}{10^{-22}  \eV}}{-1}
    \power{\frac{\velc}{200  \kms}}{-2} .
\end{align}
In this result, we have evaluated the Coulomb logarithm at $\bmax=\lsig$; thus,
$\log\Lambda_\mathrm{FDM} = \log 2\bmax/\lbarsig=\log(4\pi) \approx 2.5$.

We remark that the term ``heating'' is misleading: the interaction with \FDM\
fluctuations adds energy to the stellar population, thereby causing it to
expand, but the velocity dispersion of the stars may either grow or decay as a
result of this expansion depending on the radial profile of the gravitational
potential of the galaxy. To illustrate this process, we followed the evolution
of a population of test particles representing stars in an isothermal density
distribution (eq.~\ref{eq:sis}). The self-consistent gravitational potential of
this distribution is $\Phi(r)=\velc^2\log r$, which we modified to
$\Phi(r)=\tfrac{1}{2}\velc^2 \log(r^2+r_0^2)$ for reasons given below. The
initial velocities of the test particles were drawn from an isotropic
Maxwellian distribution with velocity dispersion $\sigma_\rt$ and the initial
positions were drawn from the radial distribution
$dn \propto r^2 {(r_0^2 + r^2)}^{-\velc^2/(2\sigma_\rt^2)}dr$, which ensures
that the initial phase-space distribution is a stationary solution of the
collisionless Boltzmann equation. We introduced a core radius $r_0$ into the
potential, so the integral over the radial distribution remains convergent when
$\sigma_{\rt}^2 \le 3 \velc^2$. The actual value of $r_0$ is unimportant since we
set $r_0= 0.2 \lsig$ and turned off the diffusion coefficients when
$r < \lsig$. We used the diffusion coefficients from
\eqs~\eqref{eq:Dpar_f}--\eqref{eq:D2pera}.

\begin{figure*}[ht]
  \includegraphics[width=0.49\textwidth]{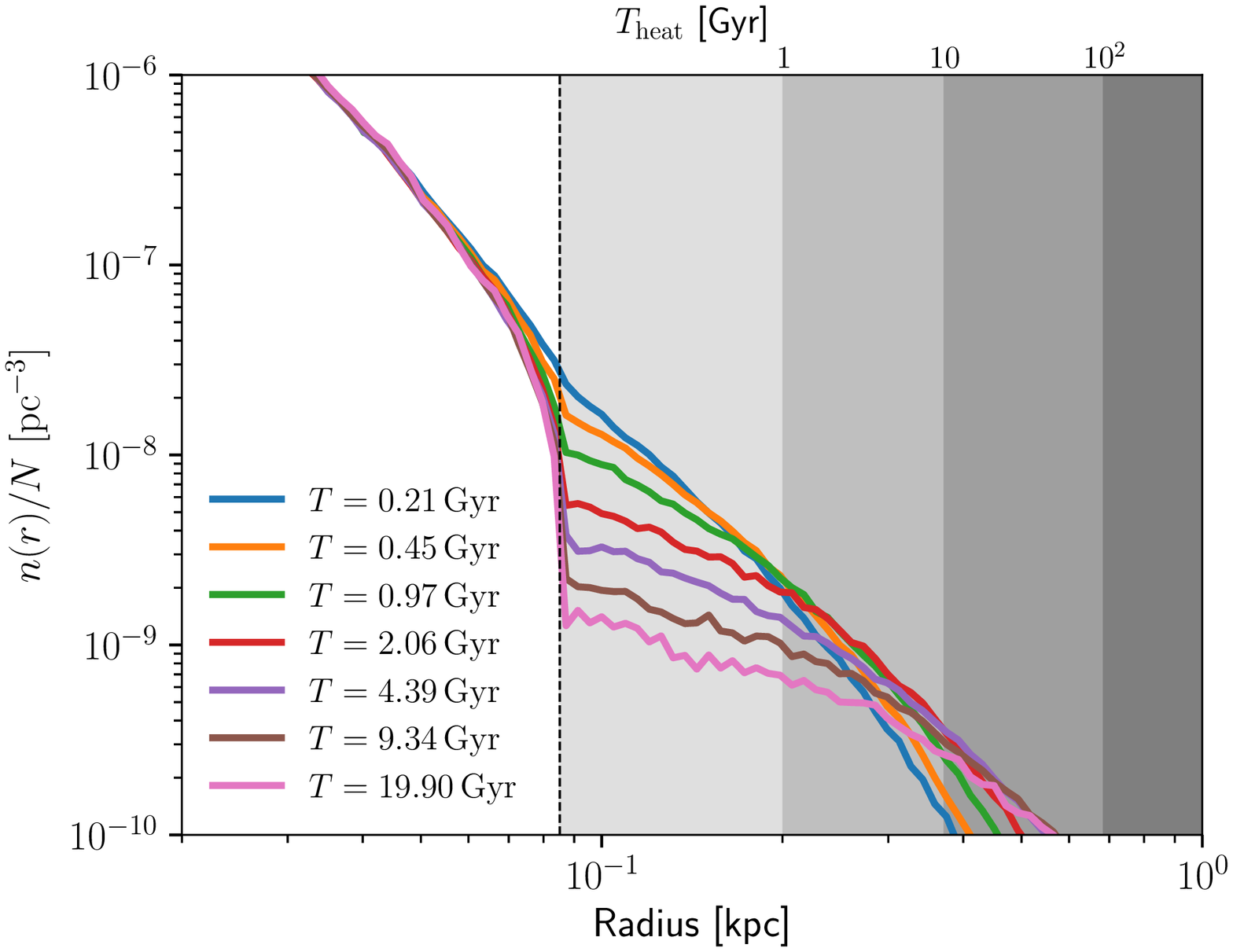}\includegraphics[width=0.49\textwidth]{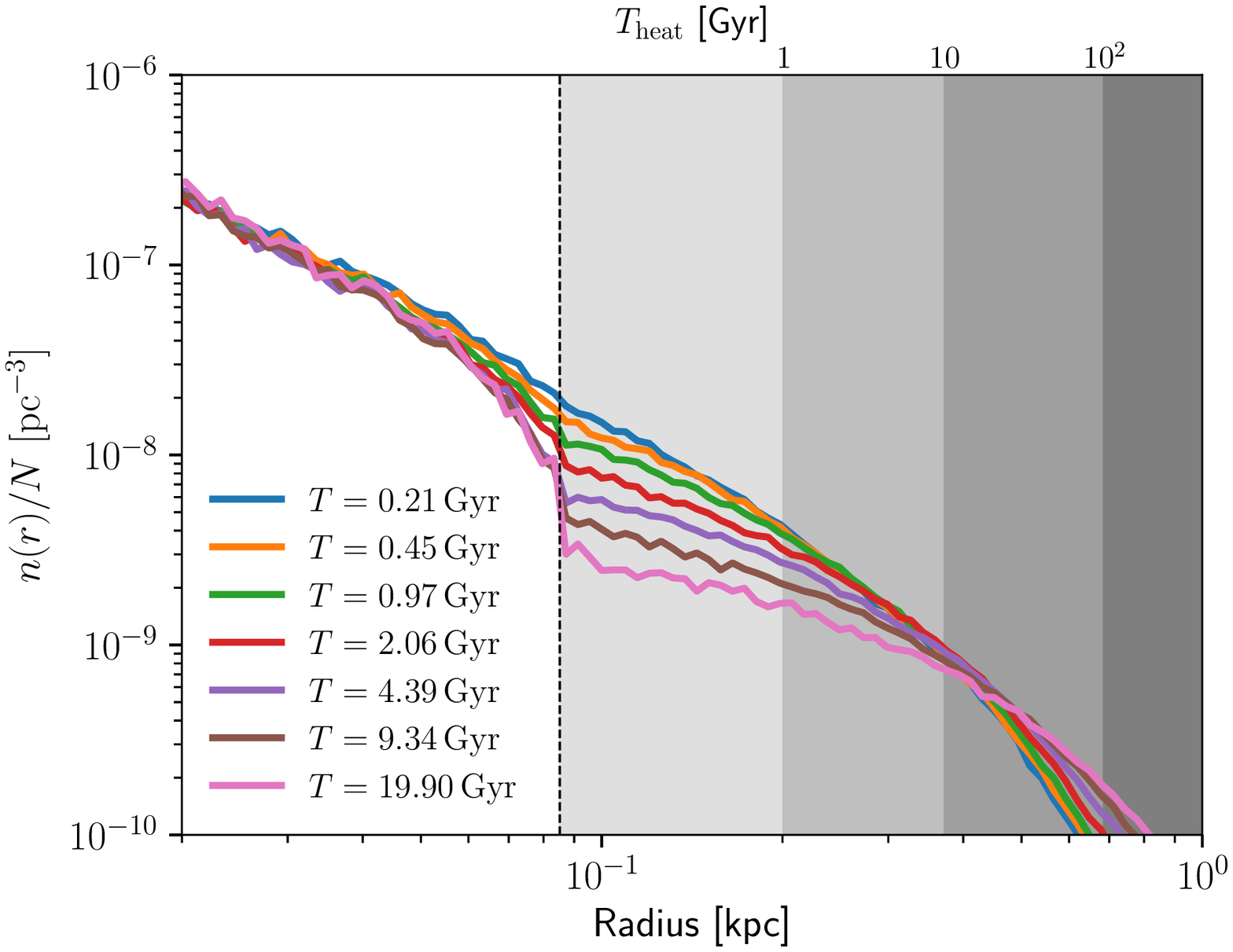}
  \includegraphics[width=0.49\textwidth]{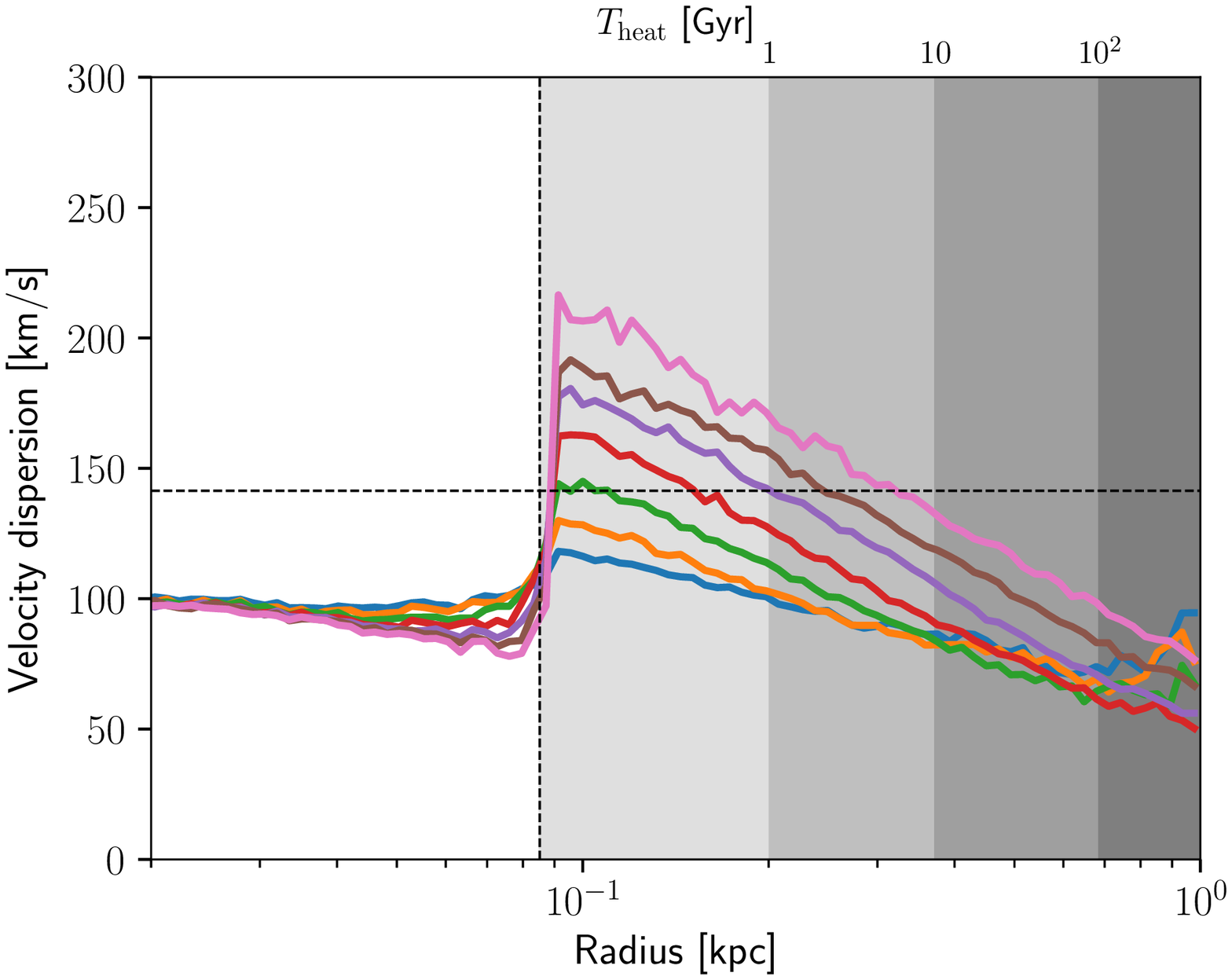}\includegraphics[width=0.49\textwidth]{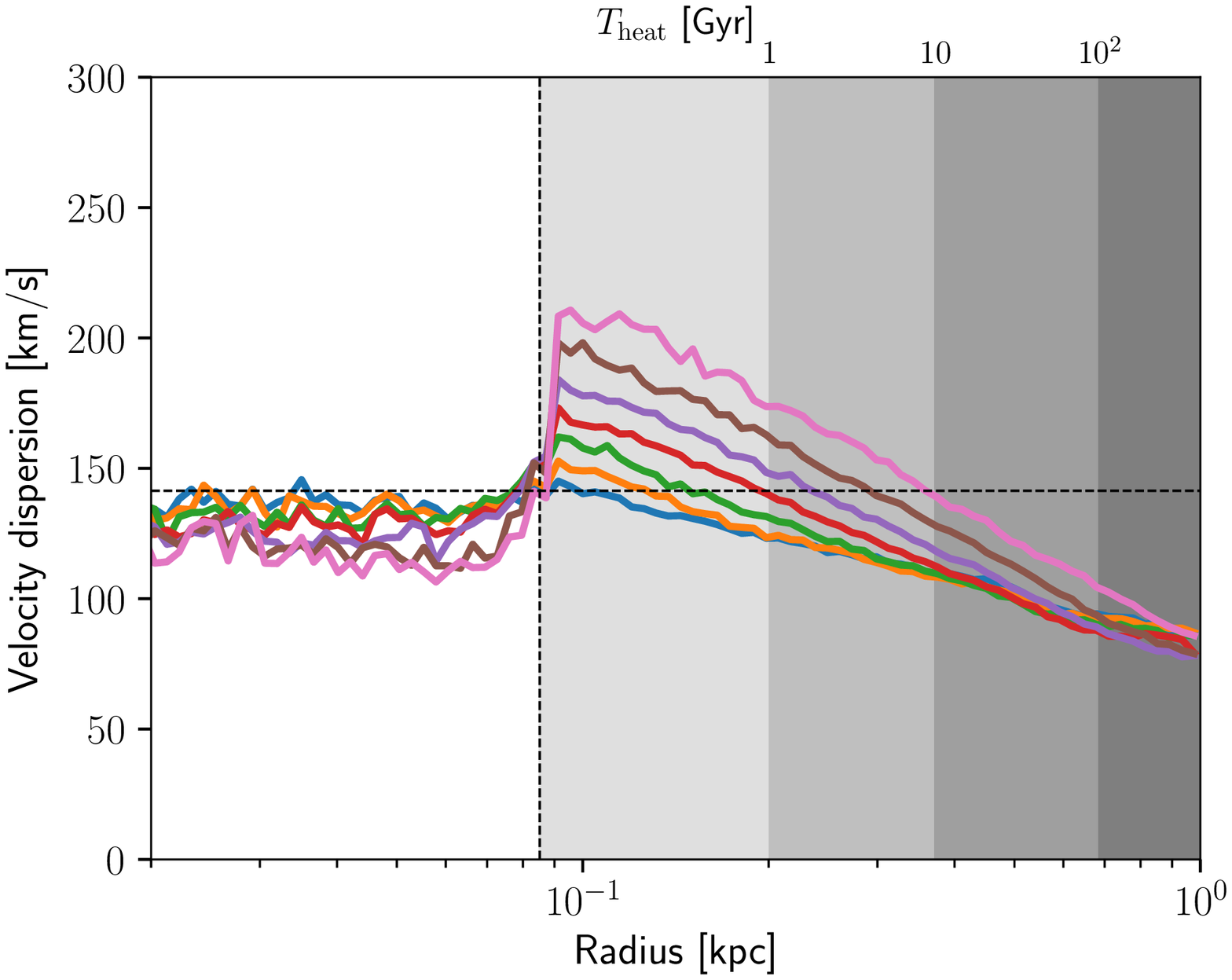}
  \caption{The expansion of a system of test particles in an isothermal \FDM\@
    halo with particle mass $\mb = 10^{-21} \eV$ and circular speed
    $\velc = 200 \kms$. The velocity dispersion
    $\sigma=\velc/\sqrt{2} \simeq 141 \kms$ (dashed horizontal lines). The
    test-particle distribution is initially isothermal with velocity dispersion
    $\sigma_\rt=\velc/2 = 100 \kms$ (left panels) and
    $\velc/\sqrt{2} \approx 141 \kms$ (right panels). In the region outside the
    de~Broglie wavelength $\lsig$ (dashed vertical lines), where the heating
    time is less than the age (top axis and shaded regions), the number density
    decreases (upper panels) and the velocity dispersion increases (bottom
    panels) as a function of time.\label{fig:num_exp}}
\end{figure*}

In Figure~\ref{fig:num_exp} we show the evolution in an \FDM\ halo having
$\velc = 200 \kms$ and $\mb = 10^{-21} \eV$. This figure shows the expansion
(upper panels) and heating (lower panels) of a system of $5600$ test
particles with initial velocity dispersion $\sigma_\rt = \velc/2$
(left panels) and $\sigma_\rt = \velc / \sqrt{2}$ (right panels). In both
cases, within a few Gyr, the velocity dispersion of the test particles exceeds
that of the \FDM\ (dashed horizontal lines), and the stellar density develops a
core in the region outside the de~Broglie wavelength and inside the radius
where $T_\mathrm{heat}$ equals the age (shaded regions). Our assumption that
fluctuations in the \FDM\ density have no effect inside the de~Broglie
wavelength $\lsig$ is oversimplifying, so the sharp changes in density
and dispersion at $\lsig$ are unrealistic.

\begin{figure}[ht]
  \plotone{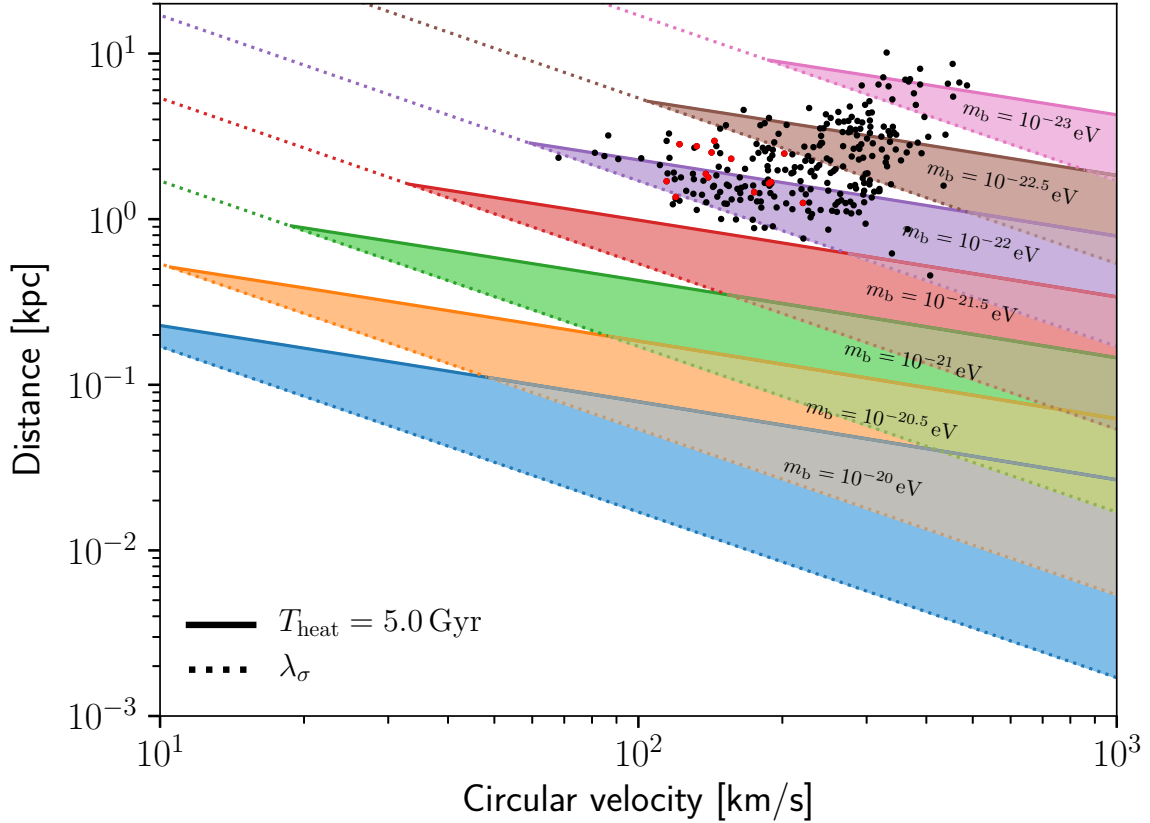}
  \caption{The heated region, in which $T_\mathrm{heat} \le 5 \Gyr$ (solid
    lines, eq.~\ref{eq:t_heat_bis}) and $r > \lsig$ (dashed lines), for several
    values of the boson mass $\mb$. In the colored regions, old stellar systems
    will be heated significantly, causing the stellar system to expand and its
    velocity dispersion to grow (see Figure~\ref{fig:num_exp}). For comparison,
    we show as circles the projected half-light radius and maximum circular
    speed $v_{\rc}^{\max}$ (circles) for the ATLAS$^\mathrm{3D}$ sample of 260
    early-type galaxies~\citep{Cappellari+2013}. The effects of \FDM\ heating
    are overestimated in most of these because they are not dark-matter
    dominated near their centers; however, the 14 galaxies marked in red are
    estimated to have a dark-matter fraction larger than
    $0.5$~\citep{Cappellari+2013b}.\label{fig:rheat}}
\end{figure}

In Figure~\ref{fig:rheat}, we show the heated region as a function of the
circular speed for several values of the particle mass $\mb$, along with the
effective radii and maximum circular speeds for the ATLAS$^\mathrm{3D}$ sample
of elliptical galaxies~\citep{Cappellari+2013}. These galaxies are most
sensitive to particle masses in the range $\mb \sim 10^{-22}$--$10^{-23} \eV$,
which is somewhat smaller than the mass range of interest for influencing
small-scale structure. To probe larger masses we need to look for evidence of
heating at smaller radii, but here, (i) most galaxies are not dark-matter
dominated, and (ii) the \FDM\ may be in the form of a ground-state soliton and
thus would not heat the stars.

\section{Summary and Conclusions}
\label{sec:summary}

Fuzzy dark matter (\FDM) is an intriguing alternative to \CDM\ that may resolve
some or all of the failures of \CDM\ to predict the properties of the structure
of galaxies on scales less than $30\kpc$ or so. \FDM\ exhibits a rich set of
novel phenomena. In particular, the density and gravitational potential
fluctuations in an isolated \CDM\ halo gradually decay as sub-halos are
destroyed and tidal tails are phase-mixed. In contrast, an isolated \FDM\ halo
exhibits persistent density fluctuations that arise because of the limited
number of eigenstates that it contains.

A test particle moving through the fluctuating \FDM\ potential is subject to
stochastic velocity changes. We calculated the diffusion coefficients that
govern its resulting orbital evolution. For a Maxwellian velocity distribution
with dispersion $\sigma$, these diffusion coefficients are the same as the
diffusion coefficients in a classical $N$-body system if (i) the classical
system is assumed to be composed of quasiparticles with an effective mass
$\meff$ that depends on the local density (eq.~\ref{eq:meff}); and (ii) the
velocity dispersion of the quasiparticles is taken to be $\sigma/\sqrt{2}$;
(iii) the lower limit of the range of scales in the classical Coulomb
logarithm, usually taken to be $b_{90}$, the impact parameter for a $90^\circ$
deflection, is replaced by $\lbarsig/2$, half of the typical
de~Broglie angular wavelength of the \FDM\@. Similarly, the dynamical friction
force on a massive particle orbiting in an \FDM\ halo is given by the classical
formula except that the Coulomb logarithm is modified as described in (iii).

In this paper, we assumed that the mean potential is infinite and
homogeneous. In the classical case, this assumption implies that the
unperturbed particles travel on straight lines at constant velocity, while in
an \FDM\ halo, it implies that the unperturbed wavefunction is a collection of
plane waves of constant wavenumber and frequency. This is a standard
simplification that is usually reasonably accurate when the Coulomb logarithm
is much larger than unity, which in turn occurs when the radial scale is much
larger than the typical wavelength. Unfortunately, the effects of \FDM\
scattering on stellar systems are typically strongest at radii that are
comparable to the wavelength. In such cases, the present derivation is
incomplete, and further numerical and theoretical studies are
needed. Nevertheless, we believe that our main conclusions are not seriously
compromised by this limitation.

We showed that a massive object that is spiraling into the center of the galaxy
by dynamical friction is subject to stochastic velocity fluctuations when it
reaches a radius where its mass is comparable to the effective mass of the
\FDM\ quasiparticles. As this point, the \FDM\ fluctuations pump energy into
the orbit at roughly the same rate that it is drained by dynamical friction, so
the inspiral will tend to stall.

Stars, which are much lighter than the \FDM\ quasiparticles, will on average
gain energy from the \FDM\ fluctuations in the region where the relaxation time
is much smaller than the age of the galaxy. Thus, stellar systems on scales of
a few hundred pc to a few kpc will expand, and the heating time within the
stellar system will be comparable to its lifetime. Therefore, one should not
observe systems in which the heating time is much shorter than the age of the
galaxy.

At the present, time neither of these physical processes offers a robust new
constraint on the mass range of possible \FDM\ particles. Other effects of
relaxation due to \FDM\ are discussed by~\citet{Hui+2017}.

Finally, in this paper, we have discussed only the effects of \FDM\
fluctuations on the orbits of classical objects such as stars and black
holes. The effects of \FDM\ fluctuations on the \FDM\ halo itself are also
important, but these must be analyzed using other
tools~\citep[e.g.,][]{Levkov+2018, Mocz+2018}.

\acknowledgments\

We thank Philip Mocz for thoughtful comments on an earlier version of this
manuscript. BB acknowledges support from the Schmidt Fellowship. JBF
acknowledges support from Program number HST-HF2--51374 which was provided by
NASA through a grant from the Space Telescope Science Institute, which is
operated by the Association of Universities for Research in Astronomy,
Incorporated, under NASA contract NAS5--26555.

\appendix
\section{The classical limit of Fuzzy Dark Matter}
\label{sec:classical_lim}

In our derivations in Section~\ref{sec:dc_fdm}, we assumed that each plane wave
has a random phase and is moving with velocity $\bv = \hbar \bk /\mb$. These
assumptions are valid if the number of particles per wave
$N \sim \rhob \lambda^3 /\mb \sim \rhob \hbar^3 / (\sigma^3 \mb^4)$ is
large. Thus, our assumptions are valid when
\begin{equation}
  \label{stat_lim}
  \mb \ll m_\rs \equiv {\bigg(\frac{\rhob \hbar^3}{\sigma^3}\bigg)}^{1/4} =
      {\bigg(\frac{\hbar^3}{2\pi G \sigma r^2}\bigg)}^{1/4}  \simeq  35  \eV
      \,\power{\frac{r}{1 \kpc}}{-1/2} \power{\frac{\sigma}{200 \kms}}{-1/4},
\end{equation}
in which we have used equation (\ref{eq:sis}).  From \eq~\eqref{eq:meff}, we can
see that the condition $\mb \ll m_\rs$ is equivalent to $\mb \ll \meff$, that is,
each quasiparticle of mass $\meff$ must contain many \FDM\ particles of mass
$\mb$.

At even larger masses, the system will behave like a classical system of free
particles of mass $\mb$. Consider a free particle at position
$x \pm \Delta x_0 $ and velocity $v \pm \Delta v_0$, where $\Delta x_0$ and
$\Delta v_0$ are the initial uncertainties in position and velocity, and
$\Delta x_0 \Delta v_0 \ge \hbar / (2 \mb)$. After a time $T$, the uncertainty
in position is
$\Delta x = \Delta x_0 + \Delta v_0 T\ge \Delta x_0 (1 + \hbar T / [2\mb
{(\Delta x_0)}^2] )$. The slowest scattering events have $T\simeq T_\rd$ where
$T_\rd=r/\sigma$ is the dynamical time and $r$ is the orbit radius. If the
particles are to behave classically, the position uncertainty cannot grow
significantly during the scattering and cannot exceed the typical distance
between particles $d = {(\mb / \rhob)}^{1/3}$, demanding that
$d \gg \Delta x_0 \gg {(\hbar T_\rd / 2 \mb)}^{1/2}$. Therefore, we can define
a critical particle mass $m_\rc$ at which $d={(\hbar T_\rd / 2 \mb)}^{1/2}$, and
the particles behave classically if
\begin{equation}
  \label{stat_lima}
  \mb \gg m_\rc \equiv \rhob^{2/5}{(\hbar T_\rd/2)}^{3/5}=\frac{1}{2}{\left(\frac{\hbar^3\sigma}{\pi^2G^2r}\right)}^{1/5}
  =1.25 \times 10^{16} \eV
  \power{\frac{r}{1 \kpc}}{-1/5}
  \power{\frac{\sigma}{200 \kms}}{1/5}.
\end{equation}
Thus, there is an intermediate range $m_\rs \lesssim \mb \lesssim m_\rc$ in
which the behavior needs further investigation, which we now undertake.

To make these arguments more quantitative we consider the following wave function: 
\begin{equation}
  \label{eq:psi_c}
  \psi(\br, t) =  \dint \bk\, 
  \varphi(\bk)
  \re^{\ri \bk \cdot \br - \ri \omega(k) t},
\end{equation}
where $\omega(k)$ is given by \eq~\eqref{eq:omega}, and the wave function in
$\bk$--space is a sum of Gaussian wavepackets,
\begin{equation}
  \label{eq:phi_c}
  \varphi(\bk) = \sum_{n=1}^N \frac{\veps^{3/2}\mb^{1/2}}{2^{3/4}\pi^{9/4}}
  \re^{-{|\bk- \mb\bv_n/\hbar |}^2\veps^2} \re^{\ri \phi_n -
    \ri \bk\cdot\br_n}.
\end{equation}
Here $\{\br_n\}$, $\{\bv_n\}$ are random positions and velocities drawn from
the \DF\ $\Fb(\bv)$, $\{\phi_n\}$ are independent random phases,
 and $\veps$ is the initial uncertainty in position. The normalization is such that $\langle |\psi|^2\rangle=\rhob$.

The correlation function of $\varphi(\bk)$ is given by  $\langle
\varphi(\bk)\varphi^*(\bk') \rangle = f_k(\bk)\delta(\bk -\bk')$, where
\begin{equation}
  \label{eq:Fkphiphi}
  f_k(\bk) = \frac{8\veps^3}{{(2\pi)}^{3/2}}\dint \bv \Fb(\bv)
  \re^{-{2\veps^2|\bk - \mb\bv/\hbar|}^2},
\end{equation}
and as expected the mean density is stationary, $\rhob  =    \langle {|\psi(\br, t)|}^2
\rangle = \dint \bk\, f_k(\bk)  = \dint \bv\, \Fb(\bv)$.

Assuming $\Fb(\bv)$ is a Maxwellian \DF\ with velocity dispersion $\sigma$, the
correlation function for the density fluctuations $\rho(\br,t) = {|\psi(\br,t)|}^2 - \rhob$ is 
\begin{align}
  \label{eq:rhorho}
  \langle \rho(\br, t) \rho(\br', t') \rangle = {} 
  &
    \frac{\rhob^2}{{\Big[1 + 
    {(1 + \alpha_1^2)}^2 {(\sigma \Delta t/\lbarsig)}^2\Big]}^{3/2}}
    \exp\left[{-\frac{{(|\Delta \br|/\lbarsig)}^2 {(1 + \alpha_1^2)}}
    {1 + {(1 + \alpha_1^2)}^2 {(\sigma \Delta t/\lbarsig)}^2}}\right]
    \nonumber \\
  &
    +              
    \frac{\mb\rhob}{{8 \pi^{3/2}  \veps^3}
    {\Big[
    1 + {(\sigma \Delta t/\sqrt{2}\veps)}^2 + \alpha_2^2/2
    \Big]}^{3/2}
    }
    \exp\left[{-\frac{{(|\Delta \br|/2\veps)}^2}
    {1 + {(\sigma \Delta t/\sqrt{2}\veps)}^2 + \alpha_2^2/2}}\right].
\end{align}
Here we have defined $\Delta \br = \br - \br'$, $\Delta t = t - t'$, $T = \sqrt{t^2+t'^2}$, and two dimensionless parameters $\alpha_1 = \lbarsig/(2\veps)$ and
$\alpha_2 = \hbar T/(2\mb\veps^2)$.

Since we require the correlations to be stationary over a dynamical time
$T_\rd \sim r/ \sigma$, we can set $\veps^2 \gg \lbarsig r \gg \lbarsig^2 $ so
$\alpha_1 \ll1$. Moreover, when $T\lesssim T_\rd$, we have
$\alpha_2< \hbar T_\rd/(2\mb\veps^2)\sim \lbarsig r/\veps^2\ll1 $. Then,
\eq~\eqref{eq:rhorho} becomes
\begin{equation}
  \label{eq:rhorho_II}
  C(\br, t) = {} 
    \frac{\rhob^2}{{\big[1 + 
     {(\sigma t/\lbarsig)}^2\big]}^{3/2}}
    \exp\bigg[{-\frac{{(r/\lbarsig)}^2}
    {1 +  {(\sigma t/\lbarsig)}^2}}\bigg]
    +              
    \frac{\mb \rhob}{{8 \pi^{3/2}  \veps^3}
    {\Big[
    1 + {(\sigma t/\sqrt{2}\veps)}^2
    \Big]}^{3/2}
    }
    \exp\left[{-\frac{{(r/2\veps)}^2}
    {1 + {(\sigma t/\sqrt{2}\veps)}^2}}\right].
\end{equation}
By comparing this result to the correlation function we obtained for \FDM\
(eq.~\ref{eq:R_fdm_m}) and for classical particles (eq.~\ref{eq:R_soft_m}), we
can see that it is the sum of the \FDM\ (first term) and the classical (second
term) limits. The diffusion coefficients for a zero-mass test particle are now
(cf.~eqs.~\ref{eq:Dpar_f}--\ref{eq:D2pera} and~\ref{eq:Dpar}--\ref{eq:D2per})
\begin{align}
  \label{eq:Dpar_cl}
  D[\Delta v_\parallel] = {} & - 4\pi G^2 \rhob \bigg[
                               \frac{\meff \log\Lambda_\mathrm{FDM}}{\seff^2}
                               \mathbb{G}(\Xeff)
                               +
                               \frac{\mb \log\Lambda_\mathrm{soft}}{\sigma^2}
                               \mathbb{G}(X) 
                               \bigg], \\
  \label{eq:D2par_cl}
  D[(\Delta v_\parallel^2)] = {} & 4 \sqrt{2} \pi G^2 \rhob \bigg[ 
                                   \frac{\meff
                                   \log\Lambda_\mathrm{FDM}}{\seff}
                                   \frac{\mathbb{G}(\Xeff)}{\Xeff}                                   
                                   + 
                                   \frac{\mb
                                   \log\Lambda_\mathrm{soft}}{\sigma}
                                   \frac{\mathbb{G}(X)}{X}                                   
                                   \bigg],
  \\
  \label{eq:D2per_cl}
  D[{(\Delta\bv_\bot)}^2] = {} & 4 \sqrt{2} \pi G^2 \rhob  \bigg[
                                 \frac{\meff
                                 \log\Lambda_\mathrm{FDM}}{\seff}
                                 \frac{\erf(\Xeff) - \mathbb{G}(\Xeff)}{\Xeff}
                                 + 
                                 \frac{\mb
                                 \log\Lambda_\mathrm{soft}}{\sigma} \frac{\erf(X) - \mathbb{G}(X)}{X}
                                 \bigg],
\end{align}
where $\Lambda_\mathrm{FDM} = 2\bmax/(\lbarsig)$,
$\Lambda_\mathrm{soft} = \bmax/\veps$, $\meff$ is given by \eq~\eqref{eq:meff},
the effective velocity dispersion is $\seff = \sigma/\sqrt{2}$,
and $X = v/\sqrt{2}\sigma$,
$\Xeff = v/\sqrt{2} \seff = v/\sigma$.

From the diffusion coefficients in
\eqs~\eqref{eq:Dpar_cl}--\eqref{eq:D2per_cl}, we can see that in the limit
$\mb \ll \meff$ ($\mb \ll m_\rs$, with $m_\rs$ defined in eq.~\ref{stat_lim}),
the \FDM\ dominates the relaxation. When $\meff \ll \mb \ll m_\rc$, with
$m_\rc$ defined in \eq~\eqref{stat_lima}, the relaxation is dominated by
the ``classical'' component, although the system is not a classical system
since the size of the wavepackets associated with each particle is larger than
the inter-particle separation, $\veps \gg d$. In this regime, the wave nature
of the particles affects the relaxation only through the Coulomb logarithm
$\Lambda_\mathrm{soft} = \bmax/\veps$. Finally, when $\mb \gg m_\rc$, the
system is in the classical limit, and $\veps$ should be interpreted as the size
of the (softened) particle.

\end{document}